\documentclass[a4paper,11pt]{article}
\pdfoutput=1 

\usepackage{jheppub} 

\usepackage[T1]{fontenc} 
\usepackage[utf8]{inputenc}
\usepackage{amsmath,amssymb,amstext,amsfonts} 

\usepackage{xcolor}
\usepackage{graphicx}
\usepackage{hyperref}
\usepackage{cancel} 
\usepackage{soul}
\usepackage{float}

\makeatletter
\renewcommand*\env@matrix[1][\arraystretch]{%
  \edef\arraystretch{#1}%
  \hskip -\arraycolsep
  \let\@ifnextchar\new@ifnextchar
  \array{*\c@MaxMatrixCols c}}
\makeatother

\makeatletter
\g@addto@macro\bfseries{\boldmath}
\makeatother

\newcommand{\diff}{\mathrm{d}}

\newcommand{\ellK}{\mathrm{K}}
\newcommand{\ellE}{\mathrm{E}}
\newcommand{\eps}{\epsilon}

\title{\boldmath 
 An algorithmic approach to finding canonical differential equations for elliptic Feynman integrals}

\author[a]{Christoph Dlapa} 
\author[b]{\hspace{-5pt}, Johannes M. Henn}
\author[b,c]{and Fabian J. Wagner} 

\affiliation[a]{Deutsches Elektronen-Synchrotron DESY, Notkestr. 85, 22607 Hamburg, Germany}
\affiliation[b]{Max-Planck-Institut f\"ur Physik, Werner-Heisenberg-Institut, D-80805 M\"unchen, Germany}
\affiliation[c]{Physik Department, James-Frank-Stra{\ss}e 1, Technische Universit\"at M\"unchen, D-85748 Garching, Germany}

\emailAdd{christoph.dlapa@desy.de}
\emailAdd{henn@mpp.mpg.de}
\emailAdd{fwagner@mpp.mpg.de}

\preprint{DESY 22-189, MPP-2022-138, TUM-HEP-1427/22}

\abstract{
In recent years, differential equations have become the method of choice to compute multi-loop Feynman integrals. Whenever they can be cast into canonical form, their solution in terms of special functions is straightforward. Recently, progress has been made in understanding the precise canonical form for Feynman integrals involving elliptic polylogarithms. In this article, we make use of an algorithmic approach that proves powerful to find canonical forms for these cases. To illustrate the method, we reproduce several known canonical forms from the literature and present examples where a canonical form is deduced for the first time. Together with this article, we also release an update for INITIAL, a publicly available \textsc{Mathematica} implementation of the algorithm.
}

\begin{document} 
\maketitle
\flushbottom

\newpage


\section{Introduction}

The computation of Feynman integrals is of immense importance in perturbative Quantum Field Theory. Not surprisingly, a number of different approaches have been developed and polished to a high degree to tackle this difficult challenge. Among these methods, differential equations have proven to be one of the most powerful and therefore most widely used approaches \cite{Kotikov:1991mg,Kotikov:1990kg,Remiddi:1997ny,Gehrmann:1999as,Gehrmann:2000zt}. In this method, one first uses integration-by-parts identities to reduce a given set of Feynman integrals to a basis of so-called master integrals $\vec{f}$. Then, one computes the derivative of the master integrals with respect to the kinematic invariants. The result is a linear combination of Feynman integrals which, due to the integration-by-parts identities, can again be written in terms of the basis $\vec{f}$. For one kinematic variable $x$, the differential equations therefore take the form
\begin{equation}\label{eqDEgeneralintro}
    \frac{\partial}{\partial x}\vec{f}(x,\eps) =A(x,\epsilon)\vec{f}(x,\eps)\,,
\end{equation}
where the coefficient matrix $A(x,\epsilon)$ is a rational function in $x$ and the parameter of dimensional regularization $\epsilon$. 

Although eq. (\ref{eqDEgeneralintro}), together with a boundary condition, fully determines $\vec{f}$, the matrix $A(x,\eps)$ is often complicated, and it is difficult to solve the differential equations analytically. (For numerical approaches, see e.g. \cite{Czakon:2020vql}).
A paradigm shift occurred with the realisation that the conjecture \cite{Henn:2013pwa,Henn:2014qga} that  (\ref{eqDEgeneralintro}) can be transformed into much simpler form, 
\begin{equation}
\label{eq:intro-canDE}
    \frac{\partial}{\partial x}\vec{g}(x,\eps)=\epsilon\tilde{A}(x)\vec{g}(x,\eps) \,.
\end{equation}
with the help of a basis transformation, $\vec{g}=T \vec{f}$, for a suitable invertible matrix $T(x,\eps)$.

The two main features of (\ref{eq:intro-canDE}) are that firstly, its RHS is proportional to $\eps$, 
and secondly, the singularity structure of $\tilde{A}(x)$ manifests the Fuchsian property of the system. 
The first feature means that
the solution to (\ref{eq:intro-canDE}) in terms of a power series in $\epsilon$ is reduced to straightforward iterated integration. The second feature restricts the class of iterated integrals, and in practice often allows one to fix the boundary constants in a simple way, see e.g. \cite{Henn:2020lye}.
For example, in the case of Feynman integrals evaluating to multiple polylogarithms (MPLs) \cite{Goncharov:1998kja,Goncharov:2001iea}---a very important class of special functions in this area of research---the canonical differential equations (\ref{eq:intro-canDE}) take the specific form
\begin{equation}
\label{eq:intro-polylog-kernels}
    \tilde{A}(x)=\sum_i\mathbf{m}_i\frac{\partial}{\partial x}\log\alpha_i(x) \,,
\end{equation}
for some set of rational functions $\alpha_i(x)$ and constant (kinematic- and $\eps$-independent) matrices $\mathbf{m}_i$.
Indeed, the form \eqref{eq:intro-canDE} together with \eqref{eq:intro-polylog-kernels} makes it manifest that 
resulting iterated integrals are multiple polylogarithms (MPLs). 

It is important to mention that the canonical differential equations can conjecturally be obtained from an analysis of the loop integrand of Feynman integrals 
\cite{Cachazo:2008vp,Arkani-Hamed:2010pyv,Henn:2013pwa,Arkani-Hamed:2014via}, prior to integration. This is closely connected to the property of uniform transcendental weight (see \cite{Henn:2020omi} and references therein). If the master integrals are chosen according to this integrand analysis, one immediately finds differential equations in a canonical form, without the need for constructing a (possibly complicated) transformation matrix $T$.
See refs. \cite{Abreu:2018aqd,Chicherin:2018old,Henn:2019swt,Henn:2020lye} for state-of-the-art applications.
Alternatively, one may first compute $A$, and then try to algorithmically construct $T$. See refs. \cite{Henn:2014qga,Lee:2014ioa,Prausa:2017ltv,Meyer:2017joq,Lee:2020zfb,Henn:2020lye,Dlapa:2020cwj,Dlapa:2022nct} for various ideas in that direction, including powerful algorithmic implementations.
The idea of a canonical form and concrete ways of obtaining it has streamlined the computation of Feynman integrals and led to significant advances in the computation of Feynman integrals, and corresponding physical applications in the last ten years.

However, with increasing loop order or increasing number of kinematic variables, there are many cases of Feynman integrals evaluate to functions beyond MPLs, and 
therefore eq. \eqref{eq:intro-polylog-kernels} needs to be generalized. 
The simplest of those cases is when the matrix $\tilde{A}(x)$ involves a single type of elliptic integral satisfying a second-order differential equation. 
The natural question to ask is what the precise form of the canonical differential equations is in the elliptic case, and beyond. This question has received significant attention in recent years. 

In the literature, several cases of differential equations in the form (\ref{eq:intro-canDE}) can be found
 \cite{Adams_2018,Bogner:2019lfa,Pogel:2022yat,Muller:2022gec}.
 These results are sometimes referred to as $\epsilon$-factorized forms. 
This terminology is certainly correct in view of eq. (\ref{eq:intro-canDE}).
In this paper, we use both the expression canonical form and $\eps$-form, almost interchangeably.
For us, the term canonical form means that we look for a simplified system of differential equations as first conjectured in \cite{Henn:2013pwa}, while $\eps$-form leaves open the possibility that further simplification to $A(x)$ may be found in the future. Indeed, settling conclusively the question of the specific form of $A(x)$ is a key open problem in this research area. 
 
This important question is intimately linked to the class of iterated integrals one expects to appear in the answer. For example, in the case of MPLs, it is clear that the integration kernels in eq. (\ref{eq:intro-polylog-kernels}) are sufficiently general to cover all cases.
 What generalizes these integration kernels for elliptic cases?  
 In the literature for elliptic multiple polylogarithms, one finds a host of different representations.
 A popular class of iterated integrals involves
  modular forms (or various generalizations thereof) as integration kernels, see e.g.~\cite{Adams:2017ejb,Broedel:2018qkq,Broedel:2018rwm,Walden:2020odh}. 
The different orders in $\epsilon$ of the corresponding basis $\vec{g}$ are then
written in terms of iterated integrals of modular forms. For certain applications and integration contours, the latter class of functions is equivalent to the \emph{elliptic} multiple polylogarithms (eMPLs) \cite{Broedel:2017kkb,Broedel:2018qkq,Adams:2016xah,Levin2007nsd,Brown2011alb}, 
which are often used for the direct integration of elliptic Feynman integrals from their parametric representation.
(Because of the importance of these two types of iterated integrals, the study of their properties, relations, analytic continuation and numerical evaluation is rapidly progressing \cite{Broedel:2018iwv,Duhr:2019rrs,Walden:2020odh}.)
However, despite these advances, a final picture for what integration kernels are needed in the canonical differential equations has not yet been established.

One approach to get insights into this is to extend the integrand analysis to the elliptic case. This has been pursued in \cite{Primo:2016ebd,Primo:2017ipr,Broedel:2018qkq,Frellesvig_2022}. In elliptic cases, after taking a certain number of residues, one encounters an elliptic curve. This implies that `leading singularities' are now not just maximal residues, but rather correspond to independent integration cycles on that elliptic curve. For example, in the case of the sunrise integral, there are two independent integration cycles, which is why the key part of the differential equation is given by a coupled two-by-two system. The maximal cut integral by definition solve the corresponding second-order equation, but going back to a first-order system leaves some freedom, i.e. additional information is needed to fix the canonical form.

In this paper, we follow a complementary approach. Leveraging information from maximal cuts of elliptic integrals, and imposing desirable properties such as Fuchsian behavior, we make an ansatz for what integration kernels the differential equations may contain, and then determine algorithmically whether a transformation to such a canonical form exists.
Our goal is to extend the algorithm of \cite{Dlapa:2020cwj} from the polylogarithmic case to the elliptic case. 
To determine the unknown coefficients, our algorithm assumes that at least one of the integrals is already pure, i.e.\ it should already be part of the canonical basis $\vec{g}$. We then require that the Picard-Fuchs equation derived from $\vec{f}$ should equal the one derived from $\vec{g}$, which gives enough constraints to determine the $\epsilon$-form and therefore the rest of the integrals in $\vec{g}$.

The paper is organized as follows: Section \ref{sec:methods} is split into two parts. In section \ref{sec:algorithm} we review the algorithm of \cite{Dlapa:2020cwj} to set notation and clarify important concepts. Section \ref{sec:anstatz} discusses how the ansatz has to be adapted to the case of elliptic Feynman integrals. Specifically, our approach will be that the integration kernels should have similar properties to the ones defined in \cite{Adams:2017ejb,Adams_2018}. For this, we also take inspiration from already known $\epsilon$-forms \cite{Adams_2018,Bogner:2019lfa,Pogel:2022yat,Muller:2022gec}. We then provide several non-trivial examples in section \ref{sec:examples}. In the first example, we reproduce the known $\epsilon$-form of the kite integral family \cite{Adams_2018}. The goal of this section is to clarify how to extract the elliptic functions from a given set of differential equations. Then, in sections \ref{sec:2LNP3P} and \ref{sec:higgs}, we present previously unknown $\epsilon$-forms for an example involving square-roots and linearly-dependent derivatives, respectively. In sections \ref{sec:gravity-integrals} and \ref{sec:banana-sec-appendix} we encounter examples where we have to adapt our ansatz to functions satisfying a third-order differential equation. Lastly, we describe our implementation in section \ref{sec:implementation} and conclude in section \ref{sec:conclusions}. Three appendices provide additional details to the material covered in the main text.

\newpage


\section{Description of the method}
\label{sec:methods}

In this section, we first review the algorithm of \cite{Dlapa:2020cwj}, which itself is based on \cite{Hoschele:2014qsa}, and then describe how it naturally extends to the elliptic case.

\subsection{Review of the algorithm}
\label{sec:algorithm}

Given a basis of $n$ master integrals $\vec{f}=(f_1,\ldots,f_n)^T$ depending only on a single scale $x$, one can use integration-by-parts identities to derive the system of differential equations
\begin{equation}
\label{eq:starting-DEs}
    \frac{\partial}{\partial x}\vec{f}=A(x,\eps)\vec{f}.
\end{equation}
The key idea of the algorithm is to assume that there is one integral, e.g.\ $f_1$, which is already part of the canonical basis $\vec{g}$ and does not require any further transformations, i.e.\ $f_1=g_1$. The first step is then to remove all dependence on the remaining integrals by transforming to the basis formed only by $f_1$ and its derivatives\footnote{We assume that all $n$ derivatives are linearly independent. See \cite{Dlapa:2020cwj,Dlapa:2022nct} and \cite{Adams:2017tga} for how to handle the case of linearly dependent derivatives.}:
\begin{equation}
\label{eq:Psi-transform}
    \begin{pmatrix}
    f_1^\prime\\
    f_1^{\prime\prime}\\
    \vdots\\
    f_1^{(n)}
    \end{pmatrix}=\Psi(x,\eps)\vec{f}\,.
\end{equation}
The transformation matrix $\Psi$ is given by
\begin{equation}
    \Psi\equiv\begin{pmatrix}
    \vec{v}_1A^{[1]}\\
    \vdots \\
    \vec{v}_1A^{[n]}
    \end{pmatrix},
\end{equation}
where
\begin{align}
    A^{[1]}&=A,\\
    A^{[n]}&=\frac{\partial}{\partial x}A^{[n-1]}+A^{[n-1]}A, \qquad\text{for } n>1,
\end{align}
and $\vec{v}_1=(1,0,\ldots,0)$.
In a second step, one can invert $\Psi$ and project eq.\ \eqref{eq:Psi-transform} on the first row to obtain an $n$-th order differential equation for $f_1$ (Picard-Fuchs equation):
\begin{equation}
    f_1+\sum_{m=1}^n b_mf_1^{(m)}=0,
\end{equation}
where the coefficients are
\begin{equation}
\label{eq:bdef}
    (b_1,\ldots,b_n)\equiv-\vec{v}_1\Psi^{-1}.
\end{equation}

Next, one repeats the above steps with the matrix $A(x,\eps)$ replaced by $B(x,\eps)$, where the latter represents our ansatz for the canonical form:
\begin{equation}
\label{eq:canonical-DEs}
    \frac{\partial}{\partial x}\vec{g}=B(x,\eps)\vec{g},\qquad B(x,\eps)=\eps\left(\sum_{i} a_i(x) \mathbf{m}_i \right),
\end{equation}
and the $\mathbf{m}_i$ are constant matrices that will be determined in the following.
In analogy with eq.\ \eqref{eq:Psi-transform}, this process involves the basis change
\begin{equation}
    \begin{pmatrix}
    g_1^\prime\\
    g_1^{\prime\prime}\\
    \vdots\\
    g_1^{(n)}
    \end{pmatrix}=\Phi(x,\eps)\vec{g},
\end{equation}
where the matrix $\Phi(x,\eps)$ can now similarly be used to derive an $n$-th order differential equation for $g_1$ with coefficients given by $-\vec{v}_1\Phi^{-1}$, c.f.~eq.~\eqref{eq:bdef}. From the assumption $f_1=g_1$ it then follows that $ \vec{v}_1\Psi^{-1}=\vec{v}_1\Phi^{-1}$ and therefore
\begin{equation}
\label{eq:mainconstraint}
    \vec{v}_1\Psi^{-1}\Phi=\vec{v}_1,
\end{equation}
which can be solved for the unknown constant matrices $\mathbf{m}_i$. The transformation $\vec{f}=T\vec{g}$ is then given by $T=\Psi^{-1}\Phi$.

\subsection{Ansatz for the canonical form}
\label{sec:anstatz}

Besides choosing a suitable initial integral, a key ingredient of our algorithm is the ansatz for the canonical differential equations~\eqref{eq:canonical-DEs}.  
In the particular case where the result can be written in terms of multiple polylogarithms (MPLs), the integration kernels take the form
\begin{equation}
\label{eq:dlog-kernels}
    a_i(x) = \frac{\partial}{\partial x} \log{\alpha_i(x)},
\end{equation}
where the $\alpha_i(x)$ are rational (algebraic) functions that can often be determined through the poles of $A(x,\eps)$ (at least in the univariate case). Note that \eqref{eq:dlog-kernels} has only simple poles. It is expected that one can always find a form of $\tilde{A}(x)$ such with at most a single pole at a given singular point. This corresponds to the Fuchsian property of Feynman integrals. For more information, see the discussion in refs. \cite{Henn:2014qga,Lee:2014ioa}.
In principle, one could relax this condition and allow for double poles and even transcendental functions to appear in the integration kernels, but this is not desirable. 
In particular, ideally the integration kernels directly lead to the desired class of iterated integrals, without further manipulation (like integration by parts or shuffle algebra).

Starting at two-loop level, there are many examples of Feynman integrals that evaluate to special functions beyond MPLs. To understand when this is the case, following \cite{Henn:2014qga},
let us imagine we have found a basis $\vec{f}$ in which the matrix $A(x,\eps)$ is analytic in $\eps$, i.e.
\begin{equation}
    A(x,\eps)=A^{(0)}(x)+A^{(1)}(x)\eps+\mathcal{O}(\eps^2),
\end{equation}
and then remove the $\eps^0$-term $A^{(0)}(x)$ through a transformation $T^{(0)}(x)$ satisfying
\begin{equation}
\label{eq:T0-eq}
    \frac{\partial}{\partial x}T^{(0)}(x)=A^{(0)}(x)T^{(0)}(x).
\end{equation}
The class of functions appearing in the solution is then expected to be a superset of the class of functions appearing in $\epsilon$-form. For example, in the polylogarithmic case, the transformation $T^{(0)}(x)$ involves only rational (or possibly algebraic) functions in $x$, as well as MPLs, from which we then conclude that \eqref{eq:dlog-kernels} is expected to be sufficient for the ansatz. 
We note that it is usually enough to apply the procedure of integrating out the $\epsilon^0$-part only to the diagonal blocks of the differential equations such that  $A^{(0)}(x)$ becomes strictly lower triangular after the transformation $T^{(0)}(x)$. Our terminology will therefore be that a certain diagonal block (also called sector in the following) introduces a certain set of functions into the ansatz.

After MPLs, the next most complicated case arises when the $T^{(0)}(x)$ for a specific sector involves a function satisfying a second-order differential equation
\begin{equation}
\label{eq:general-2nd-order-DE}
    \left[\frac{\partial^2}{\partial x^2}+\alpha_1(x)\frac{\partial}{\partial x}+\alpha_0(x)\right]\Psi_{1,2}=0,
\end{equation}
where $\alpha_0(x)$ and $\alpha_1(x)$ are rational functions, see section \ref{sec:examples} for explicit examples. In principle, our ansatz then has to include all linearly independent functions which are rational in $x$, $\Psi_{1},\Psi_2$ and the derivatives $\Psi_{1}^\prime$ and $\Psi_{2}^\prime$:
\begin{equation}
    a_{k_1,\ldots,k_m}(x)=\frac{x^{k_1}\Psi_1^{k_2}\Psi_2^{k_3}(\Psi_1^\prime)^{k_4}(\Psi_2^\prime)^{k_5}}{\prod_{l=6}^m (x-c_l)^{k_l}},
\end{equation}
where the powers $k_j$ are, a priory, arbitrary integers and $c_l$ are the same singular points as in the differential equations matrix $A(x,\epsilon)$. To restrict this further, we again require that the integration kernels directly lead to a well-understood class of iterated integrals, namely the ones discussed in \cite{Adams:2017ejb}, see also \cite{Broedel:2018rwm}. {Note that this assumption has been observed to be correct for a  large class of Feynman integrals involving complete elliptic integrals, however, in general we do not expect it to hold for all such Feynman integrals. In particular, we will discuss an example violating this assumption in section \ref{sec:banana-sec-appendix}. 

Imposing this restriction leads to the following conditions on the integration kernels:
\begin{itemize}
    \item No double poles are allowed.
    \item Only one of the two solutions to \eqref{eq:general-2nd-order-DE} can appear, not both at the same time. For concreteness, we call this solution $\Psi_1$.
    \item Only $\Psi_1$ and not its derivative can appear.\footnote{We tacitly assume that \eqref{eq:general-2nd-order-DE} stems from the Picard-Fuchs equation for the scalar integral of the considered sector in some integer dimension. See also section \ref{sec:sunrise-ansatz}.}
    \item The minimum degree of $\Psi_1$ is ${-2}$.
\end{itemize}
While the first restriction reflects the well-understood Fuchsian property of the system, the remaining three conditions are imposed so that the integration kernels transform in a specific way under modular transformations, see e.g.~\cite{Adams:2017ejb,Broedel:2018rwm}.

Our ansatz is therefore
\begin{equation}
\label{eq:psi1-ansatz}
    a_{k_1,\ldots,k_m}(x)=\frac{x^{k_1}\Psi_1^{k_2}}{\prod_{l=3}^m (x-c_l)^{k_l}},\qquad k_2\geq -2,
\end{equation}
subject to the condition that there are no double poles and we only take linearly independent combinations.
Further, we expect that the maximum degree of $\Psi_1$ depends on the specific example and therefore we will come back to this point in the next section.

We note that, for the purpose of our algorithm, it is not required to have an explicit expression for the function $\Psi_1$, but it suffices to know the second-order differential equation it satisfies and to study its behavior near all singular limits such that one can restrict the ansatz to be free of double poles. This behavior can conveniently be obtained by either applying the method of Frobenius on \eqref{eq:general-2nd-order-DE} or the method of Wasow \cite{wasow1965asymptotic} on \eqref{eq:T0-eq} (see also \cite{Bruser:2018jnc}).
Lastly, we mention that the restrictions discussed in this section are sufficient for finding an $\eps$-form of nearly all examples discussed in this paper after adjusting the singular points and the function $\Psi_1$ accordingly (with the banana integral family of section \ref{sec:banana-sec-appendix} being the only exception).


\section{Examples and applications}
\label{sec:examples}

As discussed in the introduction, we show the application of our algorithm to several non-trivial examples, where each of them introduces a specific new concept. As a warm-up, we first reproduce the known $\epsilon$-form of the kite integral family. Then we discuss two new examples involving square-roots and linearly dependent derivatives, respectively. Lastly, we show on two examples the applicability of our algorithm to functions satisfying a third-order Picard-Fuchs equation.

\subsection{Kite integral family}
\label{sec:kite}

\begin{figure}[ht]
\centering
	\begin{center}
	\includegraphics[width=0.4\columnwidth]{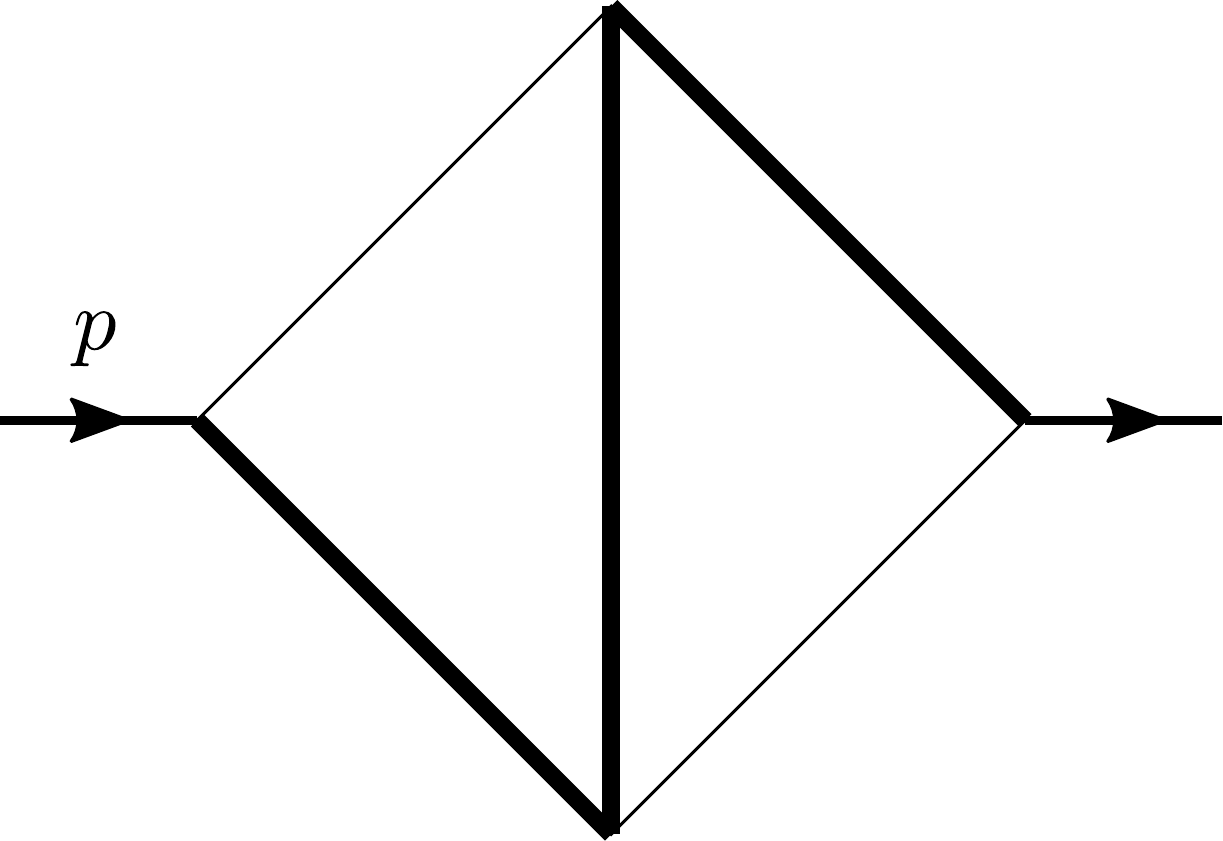}
	\end{center}
\caption{The kite integral family. Thick lines denote massive propagators. If one pinches the two massless lines, one recovers the sunrise graph which is the source of elliptic integrals.}
\label{fig:kite}
\end{figure}

The first application we want to discuss is the kite integral family shown in figure \ref{fig:kite}. One of its subsectors is the equal-mass sunrise graph, whose $\eps^0$-part is known to involve complete elliptic integrals. In ref.~\cite{Adams_2018}, it was shown that there exists a choice of master integrals which fulfill differential equations in $\eps$-form. The goal of this section is to extract the information about the elliptic functions from the differential equations matrix $A(x,\epsilon)$ and then reproduce the $\eps$-form using our algorithm. 

\subsubsection{Definitions and differential equations}
We define the kite integral family in $D$ space-time dimensions and the equal-mass case as
\begin{equation}
\label{eq:equalmasssunrisegraph}
    G_{a_1,a_2,a_3,a_4,a_5} = \int \frac{\diff^{D}k_1}{i \pi^{\frac{D}{2}}} \ \frac{\diff^{D}k_2}{i \pi^{\frac{D}{2}}} \prod_{j=1}^5 \frac{1}{D_j^{a_j}},
\end{equation}
with
\begin{equation}
    \begin{aligned}
    D_1&=-k_1^2+m^2, & D_2&=-k_2^2, & D_3&=-(k_1-k_2)^2+m^2, \\
    D_4&=-(k_1-p)^2, & D_5&=-(k_2-p)^2+m^2.
    \end{aligned}
\end{equation}
In the following, we set $m^2 \equiv 1$ for simplicity as the dependence on this mass scale can be recovered from dimensional analysis. The integrals of the kite family then only depend on the kinematic variable
\begin{equation}
	x = p^2 .
\end{equation} 
Further, we introduce the dimensional regulator $\eps$ by setting $D=D_0-2\eps$ with $D_0$ being an even integer. In particular, we will be interested in the case $D_0=4$.

Using FIRE6 \cite{Smirnov:2019qkx} and \textsc{LiteRed} \cite{Lee:2013mka}, we find that there are a total of eight master integrals, which we choose as
\begin{equation}
\begin{aligned}
\label{eq:kite-basis}
    f_1&=G_{0,0,1,0,1},\\
    f_2&=G_{0,1,1,0,1},\\
    f_3&=G_{0,1,1,1,0},\quad  f_4=G_{0,1,1,2,0},\\
    f_5&=G_{1,0,1,0,1},\quad  f_6=G_{1,0,1,0,2}\\
    f_7&=G_{1,1,0,1,1}\\
    f_8&=x\,G_{1,1,1,1,1}.
\end{aligned}
\end{equation}
Note that the normalization of $f_8$ was chosen s.t.\ its integrand has a dlog-representation with constant leading singularity in $D=4$ \cite{Cachazo:2008vp,Arkani-Hamed:2010pyv,Arkani-Hamed:2014via}. Integrals of this type are expected to evaluate to pure functions \cite{Henn:2013pwa,Arkani-Hamed:2010pyv}, which makes $f_8$ a suitable initial integral for our algorithm.

\subsubsection{Ansatz for the \texorpdfstring{$\eps$}{eps}-form}
\label{sec:sunrise-ansatz}

To construct an ansatz according to \eqref{eq:canonical-DEs}, we first analyze the differential equations for the basis given in \eqref{eq:kite-basis}. We find that they have poles in $x$ at $0, 1, 9$ and $\infty$. Therefore we also restrict the integration kernels $a_{k_1,\ldots,k_m}(x)\,\diff x$ to have poles at these points only.

To get information on the class of functions that can appear in $\eps$-form, we follow the procedure described in section \ref{sec:anstatz} and solve the differential equations at $\eps=0$ in each of the sectors. We find that only the sunrise sector formed by $f_5$ and $f_6$ requires functions beyond rational and polylogarithmic functions. In particular, we find
\begin{equation}
\label{eq:sunrise-eps0DE}
    \frac{\partial}{\partial x}\begin{pmatrix}
    f_5\\
    f_6
    \end{pmatrix}=\left[\begin{pmatrix}
    -\frac{1}{x} & \frac{3}{x} \\
    -\frac{3-x}{(x-9)(x-1)x} & \frac{9-x^2}{(x-9)(x-1)x}
    \end{pmatrix}+\mathcal{O}(D-2)\right]\begin{pmatrix}
    f_5\\
    f_6
    \end{pmatrix}+\ldots,
\end{equation}
where the ellipsis indicate terms from subsectors. Note that, here we follow \cite{Adams_2018} and perform this analysis in $D=2$ instead of $D=4$ space-time dimensions. Since the integrals in different space-time dimensions can be related through dimensional recurrence relations \cite{Tarasov_1996,Lee_2010}, this will not change the information we are trying to extract, namely the required class of functions for the ansatz.

Eq.~\eqref{eq:sunrise-eps0DE} can equivalently be written as the following second-order differential equation:
\begin{equation}
\label{eq:sunrise-PF}
    \left[\frac{\partial^2}{\partial x^2}+\frac{\left(3 x^2-20 x+9\right)}{(x-9) (x-1) x}\frac{\partial}{\partial x}+\frac{(x-3)}{(x-9) (x-1) x}\right]\Psi_{1,2}(x)=0,
\end{equation}
where $\Psi_{1,2}(x)$ indicate the homogeneous solutions for the first line in \eqref{eq:sunrise-eps0DE}. A standard choice for the solutions is given in appendix \ref{sec:ell-sunrise}. 

In summary, solving eq.~\eqref{eq:sunrise-eps0DE} requires the introduction of a function which satisfies eq.~\eqref{eq:sunrise-PF} and therefore we need to include it when constructing the ansatz for the $\eps$-form also in $D=4-2\eps$ dimensions. Further following section \ref{sec:anstatz}, we take our ansatz to consist of the integration kernels described in eq.~\eqref{eq:psi1-ansatz}, subject to the condition that they are linearly independent and free of double poles. For the latter condition it is important to know the behavior of $\Psi_1$ near all singular points, see appendix \ref{sec:banana-expansions}.
In addition, from looking at known $\eps$-forms of univariate elliptic Feynman integral families in the literature (see also the other examples in this paper), we expect that a maximum degree of two in $\Psi_1$ should be sufficient for the kite integral family.

\subsubsection{The resulting \texorpdfstring{$\epsilon$}{eps}-form}

Using the initial integral $f_8=g_8$ together with the ansatz discussed in the last section is sufficient input for our algorithm to reproduce the $\epsilon$-form first found in \cite{Adams_2018}:
\begin{equation}
\label{eq:canonicalform-kite}
    \frac{\partial}{\partial x}\vec{g}= \sum_{i,j}a_{i,j}(x)\mathbf{m}_{i,j}\,\vec{g},
\end{equation}
where the integration kernels $a_{i,j}$ are given by the set
\begin{equation}
\label{eq:kernels-kite}
    \begin{aligned}
    a_{0,1}&=\frac{1}{x(x-1)(x-9) \Psi_1^2},\\
    a_{2,1}&=\frac{1}{x}, &a_{2,2}&=\frac{1}{x-1}, &a_{2,3}&=\frac{1}{x-9},\\
    a_{3,1}&=\Psi_1, &a_{3,2}&=\frac{(x-9)\Psi_1}{x-1},\\
    a_{4,1}&=\frac{(x+3)^4 \Psi_1^2}{x(x-1)(x-9)}.
    \end{aligned}
\end{equation}
Note that this is only a subset of all kernels used in the ansatz, meaning that some of the constant coefficient matrices $\mathbf{m}_{i,j}$ were determined to be vanishing by our algorithm.
As noted before, the transformation to the basis $\vec{g}$ can be downloaded together with our public implementation, see section \ref{sec:implementation}.


\subsection{Two-loop non-planar triangle integral with internal masses}
\label{sec:2LNP3P}

\begin{figure}[ht]
\centering
	\begin{center}
	\includegraphics[width=0.4\columnwidth]{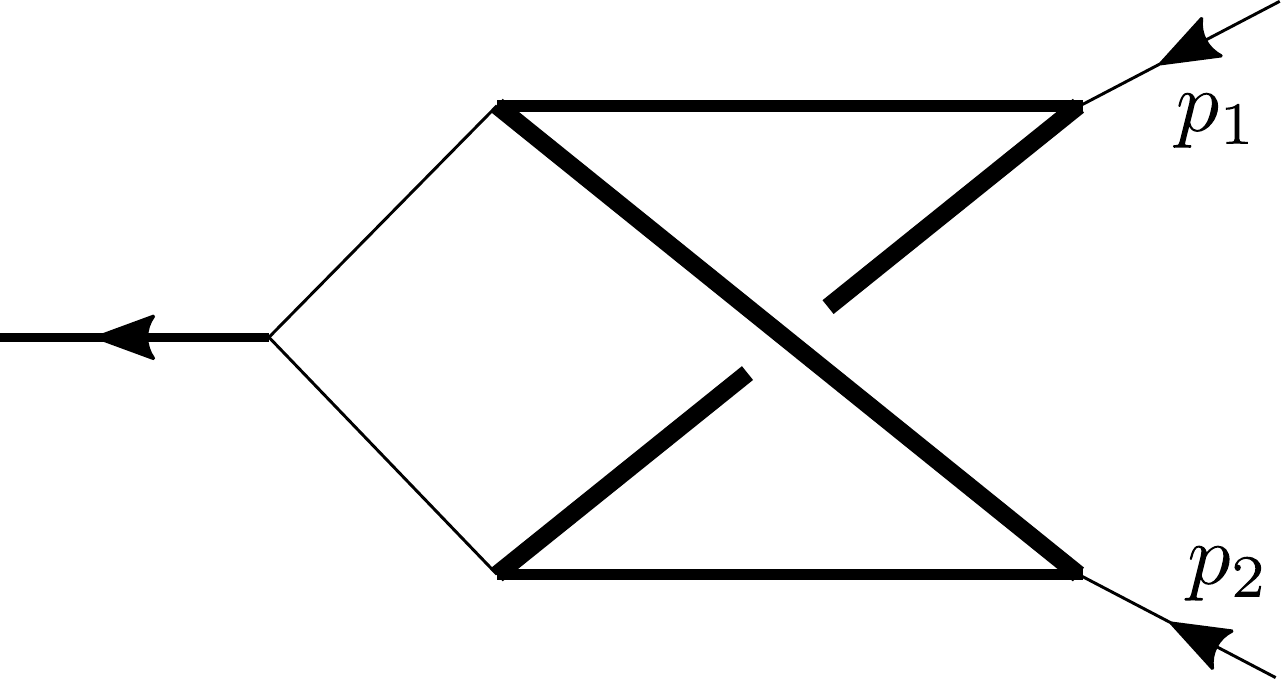}
	\end{center}
\caption{A non-planar two-loop family giving rise to elliptic integrals. The thick lines denote massive propagators whose masses are taken to be equal.}
\label{fig:np2l3p}
\end{figure}

As a second application, we consider the integral family defined by the non-planar two-loop three-point graph depicted in figure \ref{fig:np2l3p}. It has been analyzed in~\cite{NonPlanarDoubleTriangle} by means of the differential equations method where it was shown that the top topology gives rise to two master integrals that cannot be written in terms of multiple polylogarithms. Instead, the expressions presented for their finite pieces involved integrals over products of complete elliptic integrals of the first kind and polylogarithmic functions. We note that an $\eps$-form for the two integrals of the top sector was found in appendix A of \cite{Frellesvig_2022}. Here, we wish to demonstrate that our algorithm, supplied with a suitable ansatz, yields an $\eps$-form for the full differential equations (including subsectors) that leads to similar iterated integrals order-by-order in $\eps$ as were found for the kite family. 

The propagators for this example are
\begin{equation}
\begin{aligned}
	D_1&=-(k_1-p_1)^2, & D_2 &= -(k_2-p_1)^2+m^2,  \\
    D_3 &= -(k_1+p_2)^2, &	D_4 &= -(k_1-k_2+p_2)^2+m^2,  \\
    D_5 &= -(k_1 - k_2)^2+m^2,  & D_6&= -k_2^2+m^2, \\
    D_7&=-k_1^2. 
\end{aligned} 		
\end{equation}
For convenience, we set $m^2\equiv -1$ from now on. The external momenta satisfy $p_1^2=p_2^2=0$ and we define the kinematic variable 
\begin{equation}
	x\equiv (p_1+p_2)^2 .
\end{equation}
The graph from figure \ref{fig:np2l3p} corresponds to the sector $G_{1,1,1,1,1,1,0}$. All integrals in this family can be expressed in terms of eleven master integrals, two of which lie in the top sector. Reformulating the homogeneous part of the top-sector differential equations in $D=4$ as a second-order differential equation for the scalar integral, we obtain 
\begin{equation}
\label{eq:npt-PF}
    \left[\frac{\partial^2}{\partial x^2}+\frac{1}{(x-16)}\frac{\partial}{\partial x}-\frac{4}{(x-16) x^2}\right]\Psi_{1,2}=0.
\end{equation}
Consequently, as for the kite family, we have to include one of the two solutions $\Psi_{1,2}(x)$ in our ansatz. For simplicity, we take the solution $\Psi_{1}(x)$.

Analyzing the diagonal blocks of the differential equations for the nine subsector master integrals, one finds that their solutions contain the two square-roots
\begin{equation}
    r_1=\sqrt{x(x-4)}, \ \ \ \ \ \ r_2=\sqrt{x(x+4)}.
\end{equation}
Hence, as was argued in section \ref{sec:anstatz}, our ansatz has to be rational not only in $\Psi_{1}(x)$, but also $r_1$ and $r_2$.\footnote{As square-roots fulfill first-order differential equations, they can be treated by our algorithm in a similar way as $\Psi_{1}(x)$. For a detailed discussion of how to deal with square-roots, we also refer to the original article~\cite{Dlapa:2020cwj}.}

Using the same constraints as for the kite family in section~\ref{sec:sunrise-ansatz}, we can now write down a finite amount of terms for the ansatz. As the initial integral, we expect that the scalar integral divided by $\Psi_1$, i.e.\
\begin{equation}
    f_1 = g_1 = \frac{1}{\Psi_1(x)} G_{1,1,1,1,1,1,0},
\end{equation}
is a good choice. This is based on the observation that in $D=4$ the maximal cut of this integral evaluates to a pure function when integrated on a particular contour~\cite{Primo_2017,Frellesvig_2022,FabianMastersthesis}.

Applying the algorithm discussed in section \ref{sec:algorithm}, one encounters a particularity of this integral family: The sector $G_{1,1,0,1,1,1,0}$  admits one master integral that decouples completely from the remaining system of differential equations, such that they split into two separate problems. In principle, this would not pose a problem for our algorithm as we could bring them to $\eps$-form simultaneously by taking a second initial integral from this sector. However, in this case, the differential equation for $G_{1,1,0,1,1,1,0}$ is already trivial. Therefore, we just discard it and proceed with the remaining ten master integrals. An $\eps$-form is then easily obtained, where the independent integration kernels are given by the set
\begin{equation}
\label{eq:kernels-triangle}
    \begin{aligned}
    a_{0,1}&=\frac{1}{(x-16) \Psi_1^2}, \\
    a_{2,1}&=\frac{1}{x}, &a_{2,2}&=\frac{1}{x-4}, &a_{2,3}&=\frac{1}{x+4}, \\
    a_{2,4}&=\frac{1}{x-16}, &a_{2,5}&=\frac{1}{\sqrt{x(x-4)}}, &a_{2,6}&=\frac{1}{\sqrt{x(x+4)}},\\
    a_{3,1}&=\frac{\Psi_1}{x}, &a_{3,2}&=\frac{\Psi_1}{\sqrt{x(x-4)}},\\
    a_{4,1}&=\frac{(x-8)^2 \Psi_1^2}{x^2(x-16)}.
    \end{aligned}
\end{equation}
This $\epsilon$-form completes the result for the nine subsector integrals in~\cite{NonPlanarDoubleTriangle} and the canonical differential equations for the maximal cut of the top sector presented in appendix A of \cite{Frellesvig_2022}, by also providing a transformation for the off-diagonal blocks. Further, this example demonstrates nicely how square-roots are handled easily by the algorithm. 

\subsection{N3LO Higgs-production phase-space integrals}
\label{sec:higgs}

\begin{figure}[ht]
\centering
	\begin{center}
	\includegraphics[width=0.4\columnwidth]{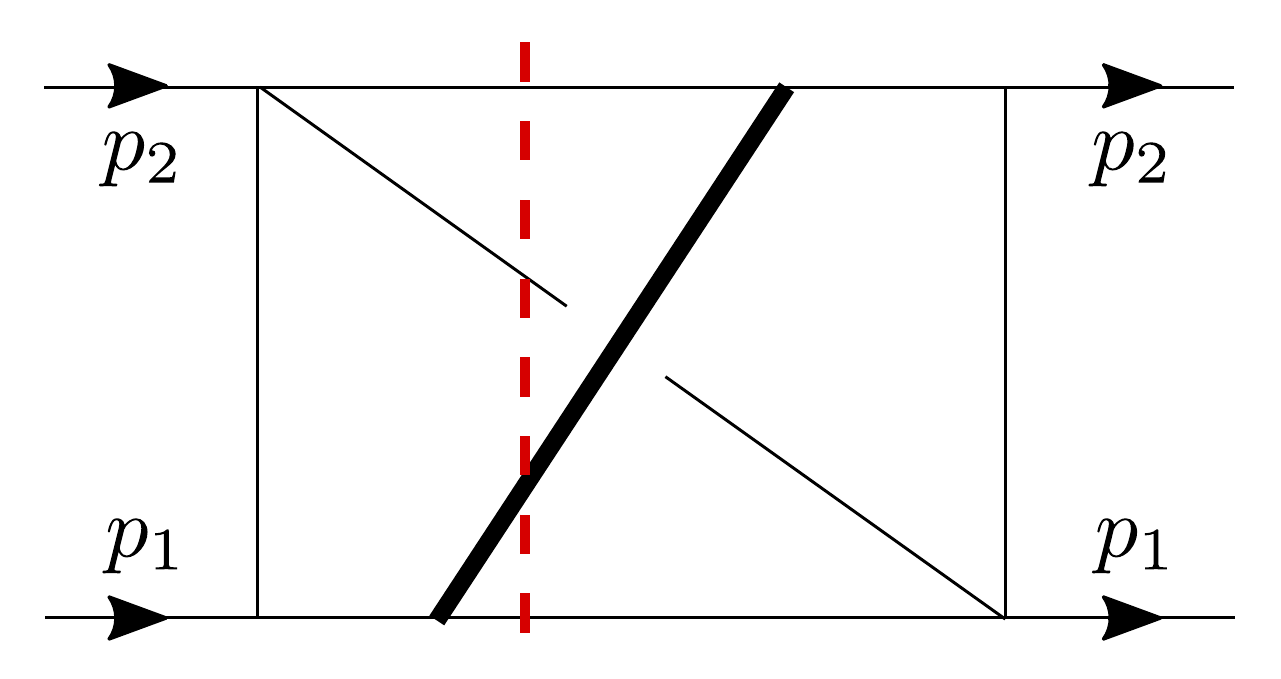}
	\end{center}
\caption{Elliptic sector in the phase-space integrals for Higgs boson production at N$^3$LO in QCD. The thick line denotes the massive Higgs line. Lines crossing the dashed line denote cut propagators.}
\label{fig:Higgs-dia-ell}
\end{figure}

As a third example, we consider some elliptic phase space integrals which appear in the calculation of the partonic coefficient functions for Higgs production at N$^3$LO in QCD~\cite{Mistlberger:2018etf}. The associated graph is depicted in figure~\ref{fig:Higgs-dia-ell} and the propagators are
\begin{equation}
\begin{aligned}
	D_1&=-(p_{12}+k_{123})^2+m_h^2, & D_2&=-k_1^2, & D_3&=-k_2^2, \\
	D_4&=-k_3^2, &D_5&=-(p_1+k_{23})^2, & D_6&=-(p_2+k_{13})^2, \\
	D_7&=-(p_{12}+k_{23})^2, & D_8&=-(p_{12}+k_{13})^2, & D_9&=-k_{12}^2, \\
	D_{10}&=-k_{13}^2, & D_{11}&=-k_{23}^2, & D_{12}&=-(p_1+k_1)^2,
\end{aligned} 		
\end{equation}
where we use the notation $p_{i_1\ldots i_n}=p_{i_1}+\ldots+p_{i_n}$.
Note that $D_1,\ldots,D_4$ are cut propagators, which means that integrals with non-positive $a_1,\ldots,a_4$ are zero. The kinematics is $s=(p_1+p_2)^2, p_1^2=p_2^2=0$ and we set $x=m_h^2$ and $s=1$.

We work in $D=4-2\eps$ space-time dimensions and consider the sector $G_{1,1,1,1,1,1,1,1,0,0,0,0}$ together with its four non-trivial subsectors. They give rise to a total of 19 master integrals. Examining the diagonal blocks of the differential equations at $\eps=0$, one finds in the top sector the second-order differential equation 
\begin{equation}
    \left[\frac{\partial^2}{\partial x^2}+\frac{\left(3 x^2-22 x-1\right) }{x \left(x^2-11
   x-1\right)}\frac{\partial}{\partial x}+\frac{(x-3) }{x \left(x^2-11 x-1\right)}\right]\Psi_{1,2}=0.
\end{equation}
In \cite{Mistlberger:2018etf} it was shown that the two independent solutions can be expressed in terms of complete elliptic integrals of the first kind. Following the same principles as before, an ansatz for the $\eps$-form including one of the two, for concreteness $\Psi_{1}$, is then written down easily and similarly to the previous section, we expect that a suitable initial integral is given by
\begin{equation}
    f_1=g_1=\frac{1}{\Psi_1}G_{1,1,1,1,1,1,1,1,0,0,0,0}.
\end{equation}
However, one finds that the first 19 derivatives of this integral are not linearly independent. Therefore, to construct an invertible $\Psi$-matrix, we need to supplement the derivatives of additional initial integrals. Through an integrand analysis or simply through trial and error, we find that already ten of the 19 basis integrals given by the IBP-reduction code \textsc{FIRE6} are uniform weight integrals and therefore this is an easy task. For more information on the initial basis $\vec{f}$, we refer to the examples that come with our implementation, see section~\ref{sec:implementation}. The independent integration kernels in the resulting $\epsilon$-form are
\begin{equation}
\label{eq:kernels-Higgs}
    \begin{aligned}
    a_{0,1}&=\frac{1}{x\left(x^2-11 x-1\right) \Psi _1^2}, &a_{2,1}&=\frac{1}{x},\\
    a_{2,2}&=\frac{1}{x+1}, &a_{2,3}&=\frac{1}{1-x},\\
    a_{2,4}&=\frac{1}{x^2-11 x-1}, &a_{2,5}&=\frac{x}{x^2-11 x-1},\\
   a_{3,1}&=\frac{(15-x) \Psi_1}{x},&a_{3,2}&=\frac{\Psi _1}{x},\\
   a_{4,1}&=\frac{\left(8 x^4+859 x^3+2062
   x^2-804 x+13\right) \Psi_1^2}{x \left(x^2-11 x-1\right)}.
    \end{aligned}
\end{equation}

\subsection{Three-loop gravitational potential integrals}
\label{sec:gravity-integrals}

\begin{figure}[H]
\centering
	\begin{center}
	\includegraphics[width=0.4\columnwidth]{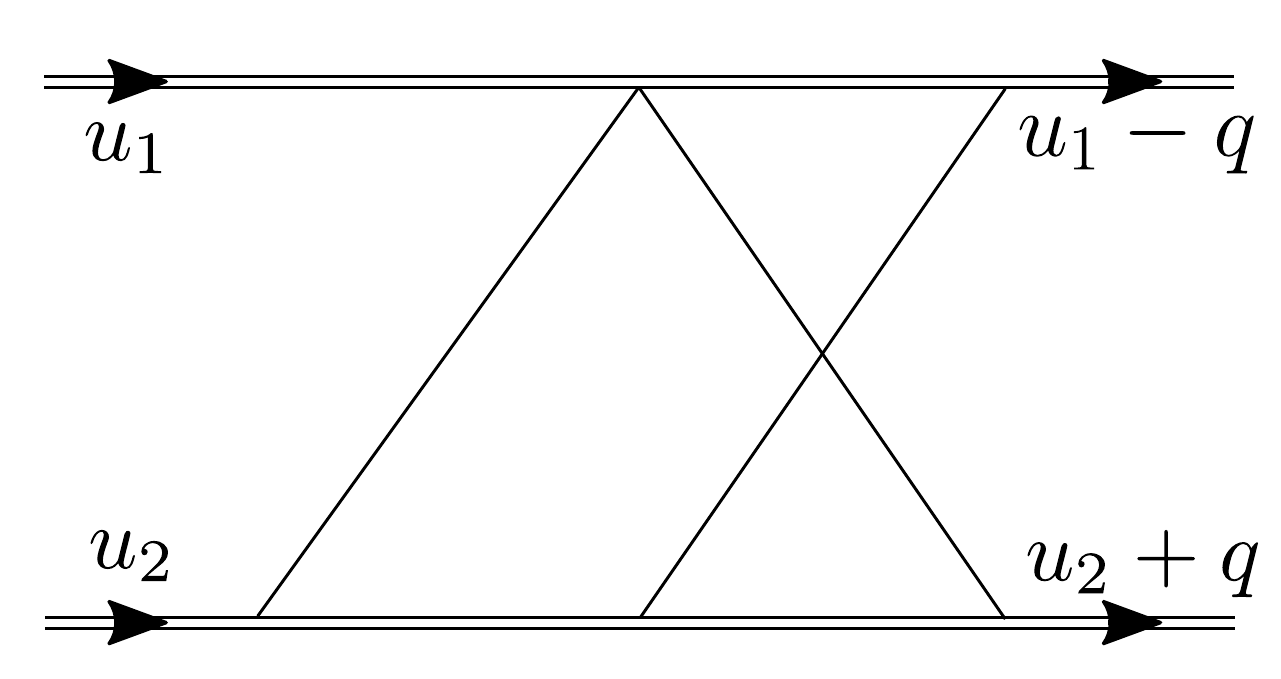}
	\end{center}
\caption{A three-loop integral sector that gives rise to elliptic functions in the gravitational potential of non-spinning binaries. Double lines denote linear propagators.}
\label{fig:Potential integrals}
\end{figure}
In this section, we discuss a family of integrals that appears in the computation of the gravitational potential of two non-spinning binaries, see e.g.~\cite{Kalin:2020mvi,Dlapa:2021npj,Dlapa:2021vgp,Dlapa:2022lmu}.
The propagators in this case are
\begin{equation}
\begin{aligned}
	D_1&=k_1\cdot u_1, & D_2&=k_2\cdot u_2, & D_3&=k_3\cdot u_2,\\
	D_4&=-k_1\cdot u_2, & D_5&=-k_2\cdot u_1, & D_6&=-k_3\cdot u_1,\\
	D_7&=-k_1^2, & D_8&=-k_2^2, & D_9&=-k_3^2, \\
	D_{10}&=-(k_1-q)^2, & D_{11}&=-(k_2-q)^2,& D_{12}&=-(k_3-q)^2,\\
	D_{13}&=-(k_1-k_2)^2, & D_{14}&=-(k_2-k_3)^2, & D_{15}&=-(k_1-k_3)^2, 
\end{aligned} 		
\end{equation}
where $D_1,D_2$ and $D_3$ are again cut propagators. The kinematics is $u_{1}^2=u_2^2=1, u_1\cdot q=u_2\cdot q=0$ and $u_1\cdot u_2=\gamma$. Further, one sets $\gamma=(x+1/x)/2$ to rationalize appearing square-roots and $q^2=-1$ since this is the only dimensionful scale and therefore the dependence of the integrals on it is trivial.

We are considering the sector $G_{1,1,1,0,0,0,0,0,1,1,1,0,0,1,1}$ depicted in figure \ref{fig:Potential integrals}, which has three master integrals. Together with its subsector, there are a total of four master integrals. Note that the differential equations for the full family, which are needed for the computation of the gravitational potential, have around 60 master integrals and have likewise been brought into $\epsilon$-form in \cite{Dlapa:2021vgp} by complementing the algorithm presented in this paper with other methods.

Inspecting the differential equations for the scalar integral in $D=4$, we find for the first time in this paper a third-order, rather than a second-order, Picard-Fuchs equation:
\begin{equation}
\label{eq:psi12-3nd-o-DE-4PMpot}
    \left[\frac{\partial^3}{\partial x^3}-\frac{6x}{1-x^2}\frac{\partial^2}{\partial x^2}+\frac{1-4x^2+7x^4}{x^2(1-x^2)^2}\frac{\partial}{\partial x}-\frac{1+x^2}{x^3(1-x^2)}\right]\tilde{\Psi}_{1,2,3}=0
\end{equation}
However, one can easily verify (see e.g.~\cite{Primo:2017ipr}) that the solutions to this equation can be written as the products
\begin{equation}
    \tilde{\Psi}_1=x\Psi_1^2,\quad \tilde{\Psi}_2=x\Psi_1\Psi_2,\quad \tilde{\Psi}_3=x\Psi_2^2,
\end{equation}
where $\Psi_{1,2}$ are the solutions of the following second-order differential equation:
\begin{equation}
\label{eq:psi12-2nd-o-DE-4PMpot}
    \left[\frac{\partial^2}{\partial x^2}+\frac{\left(1-3x^2\right) }{x(1-x^2)}\frac{\partial}{\partial x}-\frac{1}{1-x^2}\right]\Psi_{1,2}=0.
\end{equation}

Similar to before, we take the scalar integral divided by $\tilde{\Psi}_1=x\Psi_1^2$ as initial integral:
\begin{equation}
    f_1=g_1=\frac{1}{x\Psi_1^2}G_{1,1,1,0,0,0,0,0,1,1,1,0,0,1,1}.
\end{equation}
However, because the Picard-Fuchs equation is of third-order, we find that the maximal power of $\Psi_1$ in the ansatz now has to be four instead of two.\footnote{In general, if $\Psi_1$ stems from an $n$-th order Picard-Fuchs equation, the maximum degree appears to be $2n-2$.}
The integration kernels in the resulting $\epsilon$-form are
\begin{equation}
\label{eq:kernels-4PMpot}
    \begin{aligned}
    a_{0,1}&=\frac{1}{x(1-x^2)\Psi_1^2},\\
    a_{2,1}&=\frac{1+x^2}{x(1-x^2)},\\
    a_{4,1}&=\frac{(1+x^2)\Psi_1^2}{x},\quad a_{4,2}=\frac{(1+110x^2+x^4)\Psi_1^2}{x(1-x^2)},\\
    a_{6,1}&=\frac{(1+x^2)(1-18x+x^2)(1+18x+x^2)\Psi_1^4}{x(1-x^2)}
    \end{aligned}
\end{equation}
We see that our algorithm is also applicable to the case of integrals that include functions satisfying higher-order Picard-Fuchs equations, as long as we can still make a reasonable guess for the ansatz.

\subsection{Three-loop equal-mass banana graph}
\label{sec:banana-sec-appendix}

\begin{figure}[ht]
\centering
	\begin{center}
	\includegraphics[width=0.4\columnwidth]{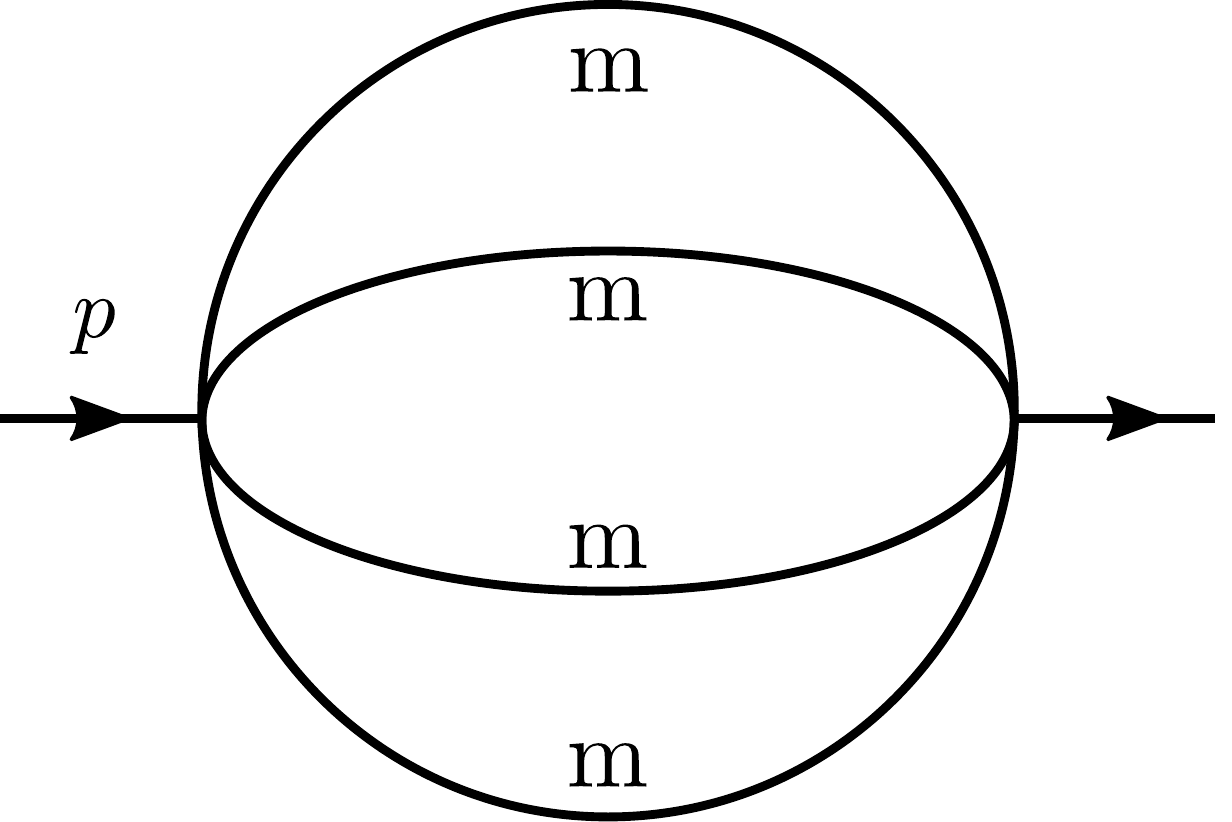}
	\end{center}
\caption{The three-loop equal-mass banana graph. The mass of the internal lines is denoted by $m$.}
\label{fig:banana-ints}
\end{figure}
As the last example, we consider the three-loop equal-mass banana graph depicted in figure \ref{fig:banana-ints}. This integral family has received a lot of attention due to being the natural extension of the sunrise graph to the next loop order. An $\eps$-form has recently been achieved in \cite{Pogel:2022yat}, see also \cite{Broedel:2019kmn,Broedel:2021zij,Primo:2017ipr}.
The propagators are
\begin{equation}
\begin{aligned}
	D_1&=-k_1^2+m^2, & D_2&=-k_2^2+m^2, & D_3&=-(k_1-k_3)^2+m^2, \\
	D_4&=-(k_2-k_3-p)^2+m^2, & D_5&=-k_3^2, & D_6&=-(k_1-p)^2, \\
	D_7&=-(k_2-p)^2, & D_8&=-(k_3-p)^2, & D_9&=-(k_1-k_2)^2,
\end{aligned} 		
\end{equation}
and we take the number of dimensions to be $D=2-2\epsilon$.
The kinematics is $p^2=-(x-1)(x-9)/x$ and $m^2=1$.
For the banana graph, we consider the three master integrals of the sector $G_{1,1,1,1,0,0,0,0,0}$ together with the single master integral from its subsector. Therefore, there are a total of four master integrals.

Setting $\eps=0$, we find the third-order differential equation
\begin{equation}
\begin{aligned}
    \bigg[\frac{\partial^3}{\partial x^3}+\frac{6(-5+x)}{(x-9)(x-1)}\frac{\partial^2}{\partial x^2}&+\frac{81-108x+156x^2-72x^3+7x^4}{(x-9)^2(x-1)^2x^2}\frac{\partial}{\partial x}\\
    &+\frac{(x-3)^3(x+3)}{(9-x)^2(x-1)^2x^3)}\bigg]\tilde{\Psi}_{1,2,3}=0
\end{aligned}
\end{equation}
for the scalar integral, similarly to the example in the previous section. Again, one can verify that the solutions can be written as
\begin{equation}
    \tilde{\Psi}_1=x\Psi_1^2,\quad \tilde{\Psi}_2=x\Psi_1\Psi_2,\quad \tilde{\Psi}_3=x\Psi_2^2,
\end{equation}
where now $\Psi_{1,2}$ are the very same functions appearing in the sunrise graph in eq.~\eqref{eq:sunrise-PF}.
We therefore take
\begin{equation}
\label{eq:banana-init-int}
    f_1=g_1=\frac{1}{x\Psi_1^2}G_{1,1,1,1,0,0,0,0,0},
\end{equation}
as initial integral and construct our ansatz accordingly. However, given our ansatz and the initial integral $f_1$, we find that there is no solution to \eqref{eq:mainconstraint}, suggesting that either \eqref{eq:banana-init-int} is not a suitable integral or that our ansatz needs to be supplemented with additional functions. Indeed, the known $\eps$-form in \cite{Pogel:2022yat} involves new integration kernels which we do not find through the procedure described in sections \ref{sec:anstatz} and \ref{sec:sunrise-ansatz}. In particular, they involve a new function, called $F_2$ in \cite{Pogel:2022yat}, which satisfies the differential equation\footnote{Here, we provide the differential equation for the product $\Psi_1 F_2$ which is somewhat simpler and for which the homogeneous part is equal to the one in \eqref{eq:sunrise-PF}. For the original differential equation for $F_2$ only, we refer to \cite{Pogel:2022yat}.}
\begin{align}
  & \hspace{-2cm}
   \Bigg[\frac{\partial^2}{\partial x^2}+\frac{\left(3 x^2-20 x+9\right)}{(x-9) (x-1) x}\frac{\partial}{\partial x}
    +\frac{(x-3)}{(x-9) (x-1) x}\Bigg]\Psi_1 F_2 \nonumber \\
& \hspace{3cm}   
=\frac{(x^2-18x+9)(x^2-2x+9)\Psi_1^3}{(x-9)(x-3)^3(x-1)x(x+3)^3},
\end{align}
After providing our algorithm with the correct ansatz taken from \cite{Pogel:2022yat}, we manage to reproduce the $\epsilon$-form.
The integration kernels are
\begin{equation}
\label{eq:kernels-banana}
    \begin{aligned}
    a_{0,1}&=\frac{6}{(x-9) (x-1) x \Psi _1^2},\\
    a_{2,1}&=\frac{x^2+9}{(x-3) x   (x+3)}, \\
    a_{2,2}&=\frac{6 F_2}{(x-9) (x-1) x \Psi _1^2},\\
   a_{4,1}&=\frac{(x-3)
   (x+3) \Psi _1^2}{12 x},\\
   a_{4,2}&=\frac{\left(x^2-18 x+9\right)^2
   \left(x^4-12 x^3+102 x^2-108 x+81\right) \Psi _1^2}{48 (x-9) (x-3)^2 (x-1)
   x (x+3)^2}-\frac{9 F_2^2}{(x-9) (x-1) x \Psi _1^2},\\
   a_{6,1}&=-\frac{F_2
   \left(x^2-18 x+9\right)^2 \left(x^4-12 x^3+102 x^2-108 x+81\right) \Psi
   _1^2}{24 (x-9) (x-3)^2 (x-1) x (x+3)^2}\\
   &+\frac{6 F_2^3}{(x-9) (x-1) x \Psi
   _1^2}+\frac{\left(x^2-18 x+9\right)^3 \left(x^2-2 x+9\right) \Psi _1^4}{36
   (x-3)^3 (x+3)^3}.
    \end{aligned}
\end{equation}
Let us comment on the singularity structure.
Although it is not obvious, all terms in eq.~\eqref{eq:kernels-banana} have at most single poles. This can be seen using the 
series expansions of $\Psi_{1,2}$ and $F_2$ given in appendix \ref{sec:banana-expansions}.

We note that one could try to enlarge the ansatz for the $\epsilon$-factorized form in a different way, in particular without the function $F_2$. Indeed, the discussion in \cite{Broedel:2021zij} suggests that integration kernels that either involve double and triple poles at certain singular points or the derivative $\Psi_1^\prime$ might be necessary. However, we  found that our algorithm did not find an $\epsilon$-factorized form based on these assumptions.


\section{Public implementation}
\label{sec:implementation}

We provide a new version of the \textsc{Mathematica} package INITIAL (an INitial InTegral ALgorithm), which is able to deal with arbitrary functions in the ansatz for the $\epsilon$-form, as long as rational replacements for their differentials w.r.t.\ the kinematic variables are also provided. As with the previous version, the package is available at
\begin{equation*}
\text{
\url{https://github.com/UT-team/INITIAL}
}
\end{equation*}
and relies on the \textsc{FiniteFlow} library \cite{Peraro:2019svx} and its dependencies. The examples mentioned in the previous section can also be downloaded from the same repository.


\section{Conclusions and outlook}
\label{sec:conclusions}

The canonical form of the differential equations has had an immense impact on our ability to compute Feynman integrals in terms of multiple polylogarithms. Beyond that, the next most complicated case is that of elliptic Feynman integrals and the goal is to advance the techniques necessary for their calculation to a similar level as for polylogarithmic Feynman integrals. In this paper, our aim was to contribute to this by extending the algorithm of \cite{Dlapa:2020cwj} to include functions satisfying higher-order Picard-Fuchs equations. In particular, we described how to extract information about these functions 
and proposed a suitable way of making an ansatz for the precise form of the canonical differential equations.
This ansatz has two main inputs: Firstly, it is motivated by known $\epsilon$-forms in the literature, and by the expected target class of iterated integrals. Secondly, as in the polylogarithmic case, it uses integrand analysis (considering multiple residues, or generalized cuts) to find a suitable `initial' integral, whose Picard-Fuchs equation is crucial input for our algorithm.
The latter then finds a complete basis of canonical integrals, assuming the ansatz for the canonical differential equations contained all relevant terms. We discussed several state-of-the-art examples, including new cases, such as the full form of the $\eps$-factorized differential equations for the massive form factor integrals in section \ref{sec:2LNP3P}, and the Higgs phase space integrals in section \ref{sec:higgs}.

Our work opens up several possible directions for further investigation:

\underline{Multivariate kinematics}: Although we focus on examples with only a single variable~$x$, the algorithm applies to multivariate cases as well, see \cite{Dlapa:2020cwj} section 3.4. A natural first example of this is the unequal-mass sunrise integral family, for which an $\epsilon$-form has been computed in \cite{Bogner:2019lfa}. The integration kernels then also involve functions satisfying an inhomogeneous second-order differential equation that can be handled by our algorithm in a similar way as $F_2$ of section \ref{sec:banana-sec-appendix}.

\underline{Two elliptic curves}: There are known examples of Feynman integral families where two of the diagonal blocks give rise to two different kinds of elliptic functions \cite{Adams:2018kez,Muller:2022gec}. The off-diagonal blocks can then simultaneously depend on both of these sectors, which complicates the process of finding an $\epsilon$-form. We are confident that our algorithm should be able to reproduce the known results \cite{Muller:2022gec}. However, it would certainly be interesting to motivate an ansatz without any information on the known $\epsilon$-form, because this should then also allow us to apply the algorithm to other problems of similar nature. 

\underline{Higher-order Picard-Fuchs equations}: We have already seen in sections \ref{sec:gravity-integrals} and \ref{sec:banana-sec-appendix} that differential equations involving a third-order differential equation at $\eps=0$ can be treated in a similar way as for second-order differential equations. However, both of these examples eventually turned out to involve elliptic functions only. It would be interesting to see how our algorithm can be applied to examples that involve functions that go beyond elliptic integrals (see e.g.~\cite{Pogel:2022ken}).

\underline{Modularity and numerical integration}: For some of our examples, the new integration kernels involve square-roots, or their denominators have algebraic roots. Therefore, it would be interesting to study their behavior under modular transformations, with the idea of relating them to functions better suited for numerical integration, see e.g.~\cite{Walden:2020odh}. 

\newpage

\acknowledgments

CD thanks Stefan Weinzierl, Ekta and Yoann Sohnle for useful discussions. JMH thanks Sebastian P\"ogel for useful discussions.
This research received funding from the European Research Council (ERC) under the European Union’s Horizon 2020 research and innovation programme,
{\it Novel structures in scattering amplitudes} (grant agreement No 725110). FW was supported by the Excellence Cluster ORIGINS funded by the Deutsche Forschungsgemeinschaft (DFG, German Research Foundation) under Germany’s Excellence Strategy - EXC-2094 - 390783311.

\appendix


\section{Elliptic functions appearing in the sunrise graph}
\label{sec:ell-sunrise}

A standard choice of solutions of \eqref{eq:sunrise-PF} is (see e.g.~\cite{Adams:2017ejb} or \cite{Adams:2018kez})
\begin{equation}
\label{eq:standard-choice}
    \Psi_1 (x) = \frac{4\ellK(k^2)}{\sqrt{Z}}, \ \ \  \Psi_2 (x) = \frac{4 i \ellK(1-k^2)}{\sqrt{Z}},
\end{equation}
with
\begin{equation}
\label{eq:auxfssunrisePsi}
    Z = (3-\sqrt{x})(1+\sqrt{x})^3, \ \ \ k^2=\frac{16 \sqrt{x}}{Z},
\end{equation}
and $\ellK$ denotes the complete elliptic integral of the first kind
\begin{equation}
    \ellK(k^2) = \int_{0}^{1} \frac{\diff z}{\sqrt{(1-z^2)(1-k^2 z^2)}}.
\end{equation}
We note that these solutions can also be obtained by integrating the maximal cut of the scalar integral over two independent integration contours~\cite{Primo_2017}.

\section{Series expansions for the kite and banana integral family}
\label{sec:banana-expansions}

The singular points of \eqref{eq:sunrise-PF} in $x$ are $0,1,9$ and $\infty$. In addition, $F_2$ also has the singular points $3$ and $-3$.
Using the method of Wasow on the $2\times2$ system or Frobenius on the second-order differential equation, the resulting solutions are ($y_1=1-x, y_9=x-9, y_\infty=1/x, y_3=x-3$ and $y_{-3}=x+3$)
\begin{align}
    \psi_{0,1}&=1+\frac{x}{3}+\frac{5 x^2}{27}+\frac{31 x^3}{243}+\mathcal{O}(x^4),\\
    \psi_{0,2}&=\frac{4 x}{9}+\frac{26 x^2}{81}+\frac{526
   x^3}{2187}+\psi_{0,1}\log(x)+\mathcal{O}(x^4),\\
    \psi_{1,1}&=1+\frac{y_1}{4}+\frac{5 y_1^2}{32}+\frac{7 y_1^3}{64}+\mathcal{O}(y_3^4),\\
    \psi_{1,2}&=\frac{3 y_1}{8}+\frac{33
   y_1^2}{128}+\frac{25 y_1^3}{128}+\psi_{1,1} \log(y_1)+\mathcal{O}(y_3^4),
\end{align}
\begin{align}
   \psi_{9,1} &=1-\frac{y_9}{12}+\frac{7 y_9^2}{864}-\frac{13 y_9^3}{15552}+\mathcal{O}(y_9^4),\\
   \psi_{9,2}&=-\frac{5
   y_9}{72}+\frac{97 y_9^2}{10368}-\frac{157
   y_9^3}{139968}+\psi_{9,1} \log(y_9)+\mathcal{O}(y_9^4),\\
   \psi_{\infty,1}&=y_\infty+3 y_\infty^2+15 y_\infty^3+93 y_\infty^4+\mathcal{O}(y_\infty^5),\\
   \psi_{\infty,2}&=4 y_\infty^2+26 y_\infty^3+\frac{526 y_\infty^4}{3}+\psi_{\infty,1}\log(y_\infty)+\mathcal{O}(y_\infty^5),\\
   \psi_{3,1}&=y_3-\frac{y_3^2}{3}+\frac{5
   y_3^3}{36}-\frac{y_3^4}{18}+\mathcal{O}(y_3^5),\\
   \psi_{3,2}&=1+\frac{y_3^3}{216}-\frac{y_3^4}{432}+\frac{y_3^5}{864}+\mathcal{O}(y_3^6),\\
   \psi_{-3,1}&=y_{-3}+\frac{y_{-3}^2}{3}+\frac{7
   y_{-3}^3}{72}+\frac{y_{-3}^4}{36}+\mathcal{O}(y_{-3}^5),\\
   \psi_{-3,2}&=1-\frac{y_{-3}^2}{48}-\frac{7
   y_{-3}^3}{864}-\frac{y_{-3}^4}{384}+\mathcal{O}(y_{-3}^5).
\end{align}
The relations of these solutions to the standard choice in \eqref{eq:standard-choice} are
\begin{align}
    \psi_{0,1}&=\frac{\sqrt{3}}{2\pi}\Psi_1,&  \psi_{0,2}&=\sqrt{3}\left(\frac{\log(3)}{\pi}\Psi_1+i \Psi_2 \right) & ,\\
    \psi_{1,1}&=-\frac{2i}{\pi}\Psi_2,&
    \psi_{1,2}&=-\frac{2}{3}\Psi_1-\frac{2i}{\pi}\log(8)\Psi_2,\\
   \psi_{9,1}&=-\frac{2\sqrt{3}}{\pi}\left(\Psi_1-\Psi_2\right),&
   \psi_{9,2}&=-\frac{2\sqrt{3}}{\pi}\left((i\pi+\log(72))\Psi_1-\log(72)\Psi_2\right),\\
   \psi_{\infty,1}&=-\frac{1}{2\pi i}\Psi_1,&
   \psi_{\infty,2}&=\frac{1}{3}\left(\Psi_1-\Psi_2\right),
\end{align}
\begin{align}
    \psi_{3,1}&=-\frac{12 \ellK\left(-7+4 \sqrt{3}\right) \Psi _1}{\sqrt{3+2
   \sqrt{3}} \pi }-\frac{12 i \ellK\left(8-4 \sqrt{3}\right) \Psi _2}{\sqrt{3+2
   \sqrt{3}} \pi },\\
       \psi_{3,2}&=\frac{\sqrt{3+2 \sqrt{3}} \Psi _1 \left(3 \ellE\left(-7+4
   \sqrt{3}\right)-2 \sqrt{3} \ellK\left(-7+4 \sqrt{3}\right)\right)}{3 \pi
   }\\
   &-\frac{i \sqrt{3+2 \sqrt{3}} \Psi _2 \left(2 \ellK\left(8-4 \sqrt{3}\right)
   \left(2 \sqrt{3} \ellK\left(-7+4 \sqrt{3}\right)-3 \ellE\left(-7+4
   \sqrt{3}\right)\right)+3 \pi \right)}{6 \pi  \ellK\left(-7+4
   \sqrt{3}\right)},\\
   \psi_{-3,1}&=\frac{(2+2 i) \sqrt{2} \sqrt[4]{3} \left(\sqrt{3}+3
   i\right) \ellK\left(\frac{1}{2}+\frac{i \sqrt{3}}{2}\right) \Psi _1}{\pi
   }\\
   &-\frac{(2-2 i) \sqrt{2} \sqrt[4]{3} \left(\sqrt{3}+3 i\right)
   \ellK\left(\frac{1}{2}-\frac{i \sqrt{3}}{2}\right) \Psi _2}{\pi },\\
   \psi_{-3,2}&=\frac{(1+i) \Psi _1 \left(3 \left(\sqrt{3}+i\right)
   \ellE\left(\frac{1}{2}+\frac{i \sqrt{3}}{2}\right)-2 \left(2 \sqrt{3}+3
   i\right) \ellK\left(\frac{1}{2}+\frac{i \sqrt{3}}{2}\right)\right)}{\sqrt{2}
   3^{3/4} \pi }\\
   &+\frac{\left(\frac{1}{2}+\frac{i}{2}\right)
   \left(\sqrt{3}+i\right) \Psi _2 \left(\ellK\left(\frac{1}{2}-\frac{i
   \sqrt{3}}{2}\right) \left(\left(\sqrt{3}-9 i\right)
   \ellK\left(\frac{1}{2}+\frac{i \sqrt{3}}{2}\right)+6 i
   \ellE\left(\frac{1}{2}+\frac{i \sqrt{3}}{2}\right)\right)-3 i \pi
   \right)}{\sqrt{2} 3^{3/4} \pi  \ellK\left(\frac{1}{2}+\frac{i
   \sqrt{3}}{2}\right)}.
\end{align}

The series expansions of $F_2$ can also be derived either by using the method of Frobenius or Wasow in combination with variation of constants, or by starting from the explicit integral representation. The result depends on which representatives are chosen for the two periods
\begin{equation}
    \begin{pmatrix}
    \psi_1\\
    \psi_2
    \end{pmatrix}=\begin{pmatrix}
    a & b\\
    c & d
    \end{pmatrix}\begin{pmatrix}
    \Psi_1\\
    \Psi_2
    \end{pmatrix}.
\end{equation}
If we always choose $\psi_1=\psi_{x_0,1}$ for every singular point $x=x_0$, the expansions are simple power series:
\begin{equation}
    \begin{aligned}
    F_2(\Psi_1\to\psi_{0,1})&=-\frac{x}{6}+\frac{x^2}{6}-\frac{5 x^3}{54}+\frac{19 x^4}{1458}+\mathcal{O}(x^{5}),\\
    F_2(\Psi_1\to\psi_{1,1})&=\frac{y_1}{6}-\frac{5 y_1^2}{24}+\frac{y_1^3}{6}-\frac{55
   y_1^4}{768}+\mathcal{O}(y_1^{5}),\\
   F_2(\Psi_1\to\psi_{9,1})&=-\frac{y_9}{6}+\frac{y_9^2}{24}-\frac{65 y_9^3}{7776}+\frac{557
   y_9^4}{373248}+\mathcal{O}(y_9^{5}),\\
   F_2(\Psi_1\to\psi_{\infty,1})&=\frac{y_{\infty }}{6}-\frac{3 y_{\infty }^2}{2}+\frac{15 y_{\infty
   }^3}{2}-\frac{19 y_{\infty }^4}{2}+\mathcal{O}(y_{\infty}^{5})
    \end{aligned}
\end{equation}
Other choices lead to very complicated expressions involving logarithms. For example
\begin{equation}
    F_2(\Psi_1\to\psi_{0,2})=\frac{x \left(24-18 \log(x)+6 \log(x)^2-\log(x)^3\right)}{6\log(x)}+\mathcal{O}(x^2).
\end{equation}
Around the singular points $x=\pm 3$ the expansions are
\begin{equation}
    \begin{aligned}
    F_2(\Psi_1\to\psi_{3,1})&=6 y_3-y_3^2+\frac{y_3^3}{3}-\frac{y_3^4}{9}+\frac{11 y_3^5}{270}+\mathcal{O}(y_3^6),\\
   F_2(\Psi_1\to\psi_{3,2})&=\frac{6}{y_3}+\frac{13 y_3}{6}-\frac{35 y_3^2}{36}+\frac{5
   y_3^3}{12}\\
   &+\log(y_3) \left(-6+2
   y_3-\frac{2 y_3^2}{3}+\frac{5 y_3^3}{18}\right)+\mathcal{O}(y_3^4),\\
   F_2(\Psi_1\to\psi_{-3,1})&=24 y_{-3}+4 y_{-3}^2+\frac{5 y_{-3}^3}{6}+\frac{7 y_{-3}^4}{36}+\frac{209
   y_{-3}^5}{4320}+\mathcal{O}(y_{-3}^6),\\
   F_2(\Psi_1\to\psi_{-3,2})&=\frac{24}{y_{-3}}+\frac{61 y_{-3}}{6}+\frac{137 y_{-3}^2}{36}+\frac{25
   y_{-3}^3}{18}\\
   &+\log(y_{-3}) \left(24+8 y_{-3}+\frac{8
   y_{-3}^2}{3}+\frac{17 y_{-3}^3}{18}\right)+\mathcal{O}(y_{-3}^4).
    \end{aligned}
\end{equation}


 \bibliographystyle{JHEP} 
 \bibliography{initial_ell} 

\providecommand{\href}[2]{#2}\begingroup\raggedright\begin{thebibliography}{10}

\bibitem{Kotikov:1991mg}
A.V.~Kotikov, \emph{{New method of massive N point Feynman diagrams
  calculation}}, \href{https://doi.org/10.1142/S0217732391003626}{\emph{Mod.
  Phys. Lett. A} {\bfseries 6} (1991) 3133}.

\bibitem{Kotikov:1990kg}
A.V.~Kotikov, \emph{{Differential equations method: New technique for massive
  Feynman diagrams calculation}},
  \href{https://doi.org/10.1016/0370-2693(91)90413-K}{\emph{Phys. Lett. B}
  {\bfseries 254} (1991) 158}.

\bibitem{Remiddi:1997ny}
E.~Remiddi, \emph{{Differential equations for Feynman graph amplitudes}},
  {\emph{Nuovo Cim. A} {\bfseries 110} (1997) 1435}
  [\href{https://arxiv.org/abs/hep-th/9711188}{{\ttfamily hep-th/9711188}}].

\bibitem{Gehrmann:1999as}
T.~Gehrmann and E.~Remiddi, \emph{{Differential equations for two loop four
  point functions}},
  \href{https://doi.org/10.1016/S0550-3213(00)00223-6}{\emph{Nucl. Phys. B}
  {\bfseries 580} (2000) 485}
  [\href{https://arxiv.org/abs/hep-ph/9912329}{{\ttfamily hep-ph/9912329}}].

\bibitem{Gehrmann:2000zt}
T.~Gehrmann and E.~Remiddi, \emph{{Two loop master integrals for gamma*
  ---\ensuremath{>} 3 jets: The Planar topologies}},
  \href{https://doi.org/10.1016/S0550-3213(01)00057-8}{\emph{Nucl. Phys. B}
  {\bfseries 601} (2001) 248}
  [\href{https://arxiv.org/abs/hep-ph/0008287}{{\ttfamily hep-ph/0008287}}].

\bibitem{Czakon:2020vql}
M.L.~Czakon and M.~Niggetiedt, \emph{{Exact quark-mass dependence of the
  Higgs-gluon form factor at three loops in QCD}},
  \href{https://doi.org/10.1007/JHEP05(2020)149}{\emph{JHEP} {\bfseries 05}
  (2020) 149} [\href{https://arxiv.org/abs/2001.03008}{{\ttfamily
  2001.03008}}].

\bibitem{Henn:2013pwa}
J.M.~Henn, \emph{{Multiloop integrals in dimensional regularization made
  simple}}, \href{https://doi.org/10.1103/PhysRevLett.110.251601}{\emph{Phys.
  Rev. Lett.} {\bfseries 110} (2013) 251601}
  [\href{https://arxiv.org/abs/1304.1806}{{\ttfamily 1304.1806}}].

\bibitem{Henn:2014qga}
J.M.~Henn, \emph{{Lectures on differential equations for Feynman integrals}},
  \href{https://doi.org/10.1088/1751-8113/48/15/153001}{\emph{J. Phys. A}
  {\bfseries 48} (2015) 153001}
  [\href{https://arxiv.org/abs/1412.2296}{{\ttfamily 1412.2296}}].

\bibitem{Henn:2020lye}
J.~Henn, B.~Mistlberger, V.A.~Smirnov and P.~Wasser, \emph{{Constructing d-log
  integrands and computing master integrals for three-loop four-particle
  scattering}}, \href{https://doi.org/10.1007/JHEP04(2020)167}{\emph{JHEP}
  {\bfseries 04} (2020) 167}
  [\href{https://arxiv.org/abs/2002.09492}{{\ttfamily 2002.09492}}].

\bibitem{Goncharov:1998kja}
A.B.~Goncharov, \emph{{Multiple polylogarithms, cyclotomy and modular
  complexes}}, \href{https://doi.org/10.4310/MRL.1998.v5.n4.a7}{\emph{Math.
  Res. Lett.} {\bfseries 5} (1998) 497}
  [\href{https://arxiv.org/abs/1105.2076}{{\ttfamily 1105.2076}}].

\bibitem{Goncharov:2001iea}
A.B.~Goncharov, \emph{{Multiple polylogarithms and mixed Tate motives}},
  \href{https://arxiv.org/abs/math/0103059}{{\ttfamily math/0103059}}.

\bibitem{Cachazo:2008vp}
F.~Cachazo, \emph{{Sharpening The Leading Singularity}},
  \href{https://arxiv.org/abs/0803.1988}{{\ttfamily 0803.1988}}.

\bibitem{Arkani-Hamed:2010pyv}
N.~Arkani-Hamed, J.L.~Bourjaily, F.~Cachazo and J.~Trnka, \emph{{Local
  Integrals for Planar Scattering Amplitudes}},
  \href{https://doi.org/10.1007/JHEP06(2012)125}{\emph{JHEP} {\bfseries 06}
  (2012) 125} [\href{https://arxiv.org/abs/1012.6032}{{\ttfamily 1012.6032}}].

\bibitem{Arkani-Hamed:2014via}
N.~Arkani-Hamed, J.L.~Bourjaily, F.~Cachazo and J.~Trnka, \emph{{Singularity
  Structure of Maximally Supersymmetric Scattering Amplitudes}},
  \href{https://doi.org/10.1103/PhysRevLett.113.261603}{\emph{Phys. Rev. Lett.}
  {\bfseries 113} (2014) 261603}
  [\href{https://arxiv.org/abs/1410.0354}{{\ttfamily 1410.0354}}].

\bibitem{Henn:2020omi}
J.M.~Henn, \emph{{What Can We Learn About QCD and Collider Physics from N=4
  Super Yang\textendash{}Mills?}},
  \href{https://doi.org/10.1146/annurev-nucl-102819-100428}{\emph{Ann. Rev.
  Nucl. Part. Sci.} {\bfseries 71} (2021) 87}
  [\href{https://arxiv.org/abs/2006.00361}{{\ttfamily 2006.00361}}].

\bibitem{Abreu:2018aqd}
S.~Abreu, L.J.~Dixon, E.~Herrmann, B.~Page and M.~Zeng, \emph{{The two-loop
  five-point amplitude in $\mathcal{N} =4$ super-Yang-Mills theory}},
  \href{https://doi.org/10.1103/PhysRevLett.122.121603}{\emph{Phys. Rev. Lett.}
  {\bfseries 122} (2019) 121603}
  [\href{https://arxiv.org/abs/1812.08941}{{\ttfamily 1812.08941}}].

\bibitem{Chicherin:2018old}
D.~Chicherin, T.~Gehrmann, J.M.~Henn, P.~Wasser, Y.~Zhang and S.~Zoia,
  \emph{{All Master Integrals for Three-Jet Production at
  Next-to-Next-to-Leading Order}},
  \href{https://doi.org/10.1103/PhysRevLett.123.041603}{\emph{Phys. Rev. Lett.}
  {\bfseries 123} (2019) 041603}
  [\href{https://arxiv.org/abs/1812.11160}{{\ttfamily 1812.11160}}].

\bibitem{Henn:2019swt}
J.M.~Henn, G.P.~Korchemsky and B.~Mistlberger, \emph{{The full four-loop cusp
  anomalous dimension in $\mathcal{N}=4$ super Yang-Mills and QCD}},
  \href{https://doi.org/10.1007/JHEP04(2020)018}{\emph{JHEP} {\bfseries 04}
  (2020) 018} [\href{https://arxiv.org/abs/1911.10174}{{\ttfamily
  1911.10174}}].

\bibitem{Lee:2014ioa}
R.N.~Lee, \emph{{Reducing differential equations for multiloop master
  integrals}}, \href{https://doi.org/10.1007/JHEP04(2015)108}{\emph{JHEP}
  {\bfseries 04} (2015) 108} [\href{https://arxiv.org/abs/1411.0911}{{\ttfamily
  1411.0911}}].

\bibitem{Prausa:2017ltv}
M.~Prausa, \emph{{epsilon: A tool to find a canonical basis of master
  integrals}}, \href{https://doi.org/10.1016/j.cpc.2017.05.026}{\emph{Comput.
  Phys. Commun.} {\bfseries 219} (2017) 361}
  [\href{https://arxiv.org/abs/1701.00725}{{\ttfamily 1701.00725}}].

\bibitem{Meyer:2017joq}
C.~Meyer, \emph{{Algorithmic transformation of multi-loop master integrals to a
  canonical basis with CANONICA}},
  \href{https://doi.org/10.1016/j.cpc.2017.09.014}{\emph{Comput. Phys. Commun.}
  {\bfseries 222} (2018) 295}
  [\href{https://arxiv.org/abs/1705.06252}{{\ttfamily 1705.06252}}].

\bibitem{Lee:2020zfb}
R.N.~Lee, \emph{{Libra: A package for transformation of differential systems
  for multiloop integrals}},
  \href{https://doi.org/10.1016/j.cpc.2021.108058}{\emph{Comput. Phys. Commun.}
  {\bfseries 267} (2021) 108058}
  [\href{https://arxiv.org/abs/2012.00279}{{\ttfamily 2012.00279}}].

\bibitem{Dlapa:2020cwj}
C.~Dlapa, J.~Henn and K.~Yan, \emph{{Deriving canonical differential equations
  for Feynman integrals from a single uniform weight integral}},
  \href{https://doi.org/10.1007/JHEP05(2020)025}{\emph{JHEP} {\bfseries 05}
  (2020) 025} [\href{https://arxiv.org/abs/2002.02340}{{\ttfamily
  2002.02340}}].

\bibitem{Dlapa:2022nct}
C.~Dlapa, \emph{{Algorithms and techniques for finding canonical differential
  equations of Feynman integrals}}, Ph.D. thesis, Munich U., 2022.
\newblock 10.5282/edoc.29769.

\bibitem{Adams_2018}
L.~Adams and S.~Weinzierl, \emph{The $\varepsilon$-form of the differential
  equations for feynman integrals in the elliptic case},
  \href{https://doi.org/10.1016/j.physletb.2018.04.002}{\emph{Physics Letters
  B} {\bfseries 781} (2018) 270}
  [\href{https://arxiv.org/abs/1802.05020}{{\ttfamily 1802.05020}}].

\bibitem{Bogner:2019lfa}
C.~Bogner, S.~M\"uller-Stach and S.~Weinzierl, \emph{{The unequal mass sunrise
  integral expressed through iterated integrals on $\overline{\mathcal
  M}_{1,3}$}},
  \href{https://doi.org/10.1016/j.nuclphysb.2020.114991}{\emph{Nucl. Phys. B}
  {\bfseries 954} (2020) 114991}
  [\href{https://arxiv.org/abs/1907.01251}{{\ttfamily 1907.01251}}].

\bibitem{Pogel:2022yat}
S.~P\"ogel, X.~Wang and S.~Weinzierl, \emph{{The three-loop equal-mass banana
  integral in \ensuremath{\varepsilon}-factorised form with meromorphic modular
  forms}}, \href{https://doi.org/10.1007/JHEP09(2022)062}{\emph{JHEP}
  {\bfseries 09} (2022) 062}
  [\href{https://arxiv.org/abs/2207.12893}{{\ttfamily 2207.12893}}].

\bibitem{Muller:2022gec}
H.~M\"uller and S.~Weinzierl, \emph{{A Feynman integral depending on two
  elliptic curves}}, \href{https://doi.org/10.1007/JHEP07(2022)101}{\emph{JHEP}
  {\bfseries 07} (2022) 101}
  [\href{https://arxiv.org/abs/2205.04818}{{\ttfamily 2205.04818}}].

\bibitem{Adams:2017ejb}
L.~Adams and S.~Weinzierl, \emph{{Feynman integrals and iterated integrals of
  modular forms}},
  \href{https://doi.org/10.4310/CNTP.2018.v12.n2.a1}{\emph{Commun. Num. Theor.
  Phys.} {\bfseries 12} (2018) 193}
  [\href{https://arxiv.org/abs/1704.08895}{{\ttfamily 1704.08895}}].

\bibitem{Broedel:2018qkq}
J.~Broedel, C.~Duhr, F.~Dulat, B.~Penante and L.~Tancredi, \emph{{Elliptic
  Feynman integrals and pure functions}},
  \href{https://doi.org/10.1007/JHEP01(2019)023}{\emph{JHEP} {\bfseries 01}
  (2019) 023} [\href{https://arxiv.org/abs/1809.10698}{{\ttfamily
  1809.10698}}].

\bibitem{Broedel:2018rwm}
J.~Broedel, C.~Duhr, F.~Dulat, B.~Penante and L.~Tancredi, \emph{{From modular
  forms to differential equations for Feynman integrals}},  in \emph{{KMPB
  Conference}: {Elliptic Integrals, Elliptic Functions and Modular Forms in
  Quantum Field Theory}}, pp.~107--131, 2019,
  \href{https://doi.org/10.1007/978-3-030-04480-0_6}{DOI}
  [\href{https://arxiv.org/abs/1807.00842}{{\ttfamily 1807.00842}}].

\bibitem{Walden:2020odh}
M.~Walden and S.~Weinzierl, \emph{{Numerical evaluation of iterated integrals
  related to elliptic Feynman integrals}},
  \href{https://doi.org/10.1016/j.cpc.2021.108020}{\emph{Comput. Phys. Commun.}
  {\bfseries 265} (2021) 108020}
  [\href{https://arxiv.org/abs/2010.05271}{{\ttfamily 2010.05271}}].

\bibitem{Broedel:2017kkb}
J.~Broedel, C.~Duhr, F.~Dulat and L.~Tancredi, \emph{{Elliptic polylogarithms
  and iterated integrals on elliptic curves. Part I: general formalism}},
  \href{https://doi.org/10.1007/JHEP05(2018)093}{\emph{JHEP} {\bfseries 05}
  (2018) 093} [\href{https://arxiv.org/abs/1712.07089}{{\ttfamily
  1712.07089}}].

\bibitem{Adams:2016xah}
L.~Adams, C.~Bogner, A.~Schweitzer and S.~Weinzierl, \emph{{The kite integral
  to all orders in terms of elliptic polylogarithms}},
  \href{https://doi.org/10.1063/1.4969060}{\emph{J. Math. Phys.} {\bfseries 57}
  (2016) 122302} [\href{https://arxiv.org/abs/1607.01571}{{\ttfamily
  1607.01571}}].

\bibitem{Levin2007nsd}
A.~Levin and G.~Racinet, \emph{Towards multiple elliptic polylogarithms},
  \href{https://arxiv.org/abs/math/0703237}{{\ttfamily math/0703237}}.

\bibitem{Brown2011alb}
F.C.S.~Brown and A.~Levin, \emph{Multiple elliptic polylogarithms},
  \href{https://arxiv.org/abs/1110.6917}{{\ttfamily 1110.6917}}.

\bibitem{Broedel:2018iwv}
J.~Broedel, C.~Duhr, F.~Dulat, B.~Penante and L.~Tancredi, \emph{{Elliptic
  symbol calculus: from elliptic polylogarithms to iterated integrals of
  Eisenstein series}},
  \href{https://doi.org/10.1007/JHEP08(2018)014}{\emph{JHEP} {\bfseries 08}
  (2018) 014} [\href{https://arxiv.org/abs/1803.10256}{{\ttfamily
  1803.10256}}].

\bibitem{Duhr:2019rrs}
C.~Duhr and L.~Tancredi, \emph{{Algorithms and tools for iterated Eisenstein
  integrals}}, \href{https://doi.org/10.1007/JHEP02(2020)105}{\emph{JHEP}
  {\bfseries 02} (2020) 105}
  [\href{https://arxiv.org/abs/1912.00077}{{\ttfamily 1912.00077}}].

\bibitem{Primo:2016ebd}
A.~Primo and L.~Tancredi, \emph{{On the maximal cut of Feynman integrals and
  the solution of their differential equations}},
  \href{https://doi.org/10.1016/j.nuclphysb.2016.12.021}{\emph{Nucl. Phys. B}
  {\bfseries 916} (2017) 94}
  [\href{https://arxiv.org/abs/1610.08397}{{\ttfamily 1610.08397}}].

\bibitem{Primo:2017ipr}
A.~Primo and L.~Tancredi, \emph{{Maximal cuts and differential equations for
  Feynman integrals. An application to the three-loop massive banana graph}},
  \href{https://doi.org/10.1016/j.nuclphysb.2017.05.018}{\emph{Nucl. Phys. B}
  {\bfseries 921} (2017) 316}
  [\href{https://arxiv.org/abs/1704.05465}{{\ttfamily 1704.05465}}].

\bibitem{Frellesvig_2022}
H.~Frellesvig, \emph{On epsilon factorized differential equations for elliptic
  feynman integrals},
  \href{https://doi.org/10.1007/jhep03(2022)079}{\emph{Journal of High Energy
  Physics} {\bfseries 2022} (2022) }
  [\href{https://arxiv.org/abs/2110.07968}{{\ttfamily 2110.07968}}].

\bibitem{Hoschele:2014qsa}
M.~H\"oschele, J.~Hoff and T.~Ueda, \emph{{Adequate bases of phase space master
  integrals for gg $\to$ h at NNLO and beyond}},
  \href{https://doi.org/10.1007/JHEP09(2014)116}{\emph{JHEP} {\bfseries 09}
  (2014) 116} [\href{https://arxiv.org/abs/1407.4049}{{\ttfamily 1407.4049}}].

\bibitem{Adams:2017tga}
L.~Adams, E.~Chaubey and S.~Weinzierl, \emph{{Simplifying Differential
  Equations for Multiscale Feynman Integrals beyond Multiple Polylogarithms}},
  \href{https://doi.org/10.1103/PhysRevLett.118.141602}{\emph{Phys. Rev. Lett.}
  {\bfseries 118} (2017) 141602}
  [\href{https://arxiv.org/abs/1702.04279}{{\ttfamily 1702.04279}}].

\bibitem{wasow1965asymptotic}
W.~Wasow, \emph{Asymptotic expansions for ordinary differential equations},
  Pure and Applied Mathematics, Vol. XIV, Interscience Publishers John Wiley \&
  Sons, Inc., New York-London-Sydney (1965).

\bibitem{Bruser:2018jnc}
R.~Br\"user, S.~Caron-Huot and J.M.~Henn, \emph{{Subleading Regge limit from a
  soft anomalous dimension}},
  \href{https://doi.org/10.1007/JHEP04(2018)047}{\emph{JHEP} {\bfseries 04}
  (2018) 047} [\href{https://arxiv.org/abs/1802.02524}{{\ttfamily
  1802.02524}}].

\bibitem{Smirnov:2019qkx}
A.V.~Smirnov and F.S.~Chuharev, \emph{{FIRE6: Feynman Integral REduction with
  Modular Arithmetic}},
  \href{https://doi.org/10.1016/j.cpc.2019.106877}{\emph{Comput. Phys. Commun.}
  {\bfseries 247} (2020) 106877}
  [\href{https://arxiv.org/abs/1901.07808}{{\ttfamily 1901.07808}}].

\bibitem{Lee:2013mka}
R.N.~Lee, \emph{{LiteRed 1.4: a powerful tool for reduction of multiloop
  integrals}}, \href{https://doi.org/10.1088/1742-6596/523/1/012059}{\emph{J.
  Phys. Conf. Ser.} {\bfseries 523} (2014) 012059}
  [\href{https://arxiv.org/abs/1310.1145}{{\ttfamily 1310.1145}}].

\bibitem{Tarasov_1996}
O.V.~Tarasov, \emph{Connection between feynman integrals having different
  values of the space-time dimension},
  \href{https://doi.org/10.1103/physrevd.54.6479}{\emph{Physical Review D}
  {\bfseries 54} (1996) 6479} [\href{https://arxiv.org/abs/9606018}{{\ttfamily
  9606018}}].

\bibitem{Lee_2010}
R.~Lee, \emph{Space{\textendash}time dimensionality as complex variable:
  Calculating loop integrals using dimensional recurrence relation and
  analytical properties with respect to},
  \href{https://doi.org/10.1016/j.nuclphysb.2009.12.025}{\emph{Nuclear Physics
  B} {\bfseries 830} (2010) 474}
  [\href{https://arxiv.org/abs/0911.0252}{{\ttfamily 0911.0252}}].

\bibitem{NonPlanarDoubleTriangle}
A.~von Manteuffel and L.~Tancredi, \emph{A non-planar two-loop three-point
  function beyond multiple polylogarithms},
  \href{https://doi.org/10.1007/jhep06(2017)127}{\emph{JHEP} {\bfseries 06}
  (2017) 127} [\href{https://arxiv.org/abs/1701.05905}{{\ttfamily
  1701.05905}}].

\bibitem{Primo_2017}
A.~Primo and L.~Tancredi, \emph{On the maximal cut of feynman integrals and the
  solution of their differential equations},
  \href{https://doi.org/10.1016/j.nuclphysb.2016.12.021}{\emph{Nuclear Physics
  B} {\bfseries 916} (2017) 94}
  [\href{https://arxiv.org/abs/1610.08397v2}{{\ttfamily 1610.08397v2}}].

\bibitem{FabianMastersthesis}
F.~Wagner, \emph{Towards a canonical form for elliptic feynman integrals},
  Master's thesis, Munich U., 2022.

\bibitem{Mistlberger:2018etf}
B.~Mistlberger, \emph{{Higgs boson production at hadron colliders at N$^{3}$LO
  in QCD}}, \href{https://doi.org/10.1007/JHEP05(2018)028}{\emph{JHEP}
  {\bfseries 05} (2018) 028}
  [\href{https://arxiv.org/abs/1802.00833}{{\ttfamily 1802.00833}}].

\bibitem{Kalin:2020mvi}
G.~K\"alin and R.A.~Porto, \emph{{Post-Minkowskian Effective Field Theory for
  Conservative Binary Dynamics}},
  \href{https://doi.org/10.1007/JHEP11(2020)106}{\emph{JHEP} {\bfseries 11}
  (2020) 106} [\href{https://arxiv.org/abs/2006.01184}{{\ttfamily
  2006.01184}}].

\bibitem{Dlapa:2021npj}
C.~Dlapa, G.~K\"alin, Z.~Liu and R.A.~Porto, \emph{{Dynamics of binary systems
  to fourth Post-Minkowskian order from the effective field theory approach}},
  \href{https://doi.org/10.1016/j.physletb.2022.137203}{\emph{Phys. Lett. B}
  {\bfseries 831} (2022) 137203}
  [\href{https://arxiv.org/abs/2106.08276}{{\ttfamily 2106.08276}}].

\bibitem{Dlapa:2021vgp}
C.~Dlapa, G.~K\"alin, Z.~Liu and R.A.~Porto, \emph{{Conservative Dynamics of
  Binary Systems at Fourth Post-Minkowskian Order in the Large-Eccentricity
  Expansion}},
  \href{https://doi.org/10.1103/PhysRevLett.128.161104}{\emph{Phys. Rev. Lett.}
  {\bfseries 128} (2022) 161104}
  [\href{https://arxiv.org/abs/2112.11296}{{\ttfamily 2112.11296}}].

\bibitem{Dlapa:2022lmu}
C.~Dlapa, G.~K\"alin, Z.~Liu, J.~Neef and R.A.~Porto, \emph{{Radiation Reaction
  and Gravitational Waves at Fourth Post-Minkowskian Order}},
  \href{https://arxiv.org/abs/2210.05541}{{\ttfamily 2210.05541}}.

\bibitem{Broedel:2019kmn}
J.~Broedel, C.~Duhr, F.~Dulat, R.~Marzucca, B.~Penante and L.~Tancredi,
  \emph{{An analytic solution for the equal-mass banana graph}},
  \href{https://doi.org/10.1007/JHEP09(2019)112}{\emph{JHEP} {\bfseries 09}
  (2019) 112} [\href{https://arxiv.org/abs/1907.03787}{{\ttfamily
  1907.03787}}].

\bibitem{Broedel:2021zij}
J.~Broedel, C.~Duhr and N.~Matthes, \emph{{Meromorphic modular forms and the
  three-loop equal-mass banana integral}},
  \href{https://doi.org/10.1007/JHEP02(2022)184}{\emph{JHEP} {\bfseries 02}
  (2022) 184} [\href{https://arxiv.org/abs/2109.15251}{{\ttfamily
  2109.15251}}].

\bibitem{Peraro:2019svx}
T.~Peraro, \emph{{FiniteFlow: multivariate functional reconstruction using
  finite fields and dataflow graphs}},
  \href{https://doi.org/10.1007/JHEP07(2019)031}{\emph{JHEP} {\bfseries 07}
  (2019) 031} [\href{https://arxiv.org/abs/1905.08019}{{\ttfamily
  1905.08019}}].

\bibitem{Adams:2018kez}
L.~Adams, E.~Chaubey and S.~Weinzierl, \emph{{Analytic results for the planar
  double box integral relevant to top-pair production with a closed top loop}},
  \href{https://doi.org/10.1007/JHEP10(2018)206}{\emph{JHEP} {\bfseries 10}
  (2018) 206} [\href{https://arxiv.org/abs/1806.04981}{{\ttfamily
  1806.04981}}].

\bibitem{Pogel:2022ken}
S.~P\"ogel, X.~Wang and S.~Weinzierl, \emph{{The $\varepsilon$-factorised
  differential equation for the four-loop equal-mass banana graph}},
  \href{https://arxiv.org/abs/2211.04292}{{\ttfamily 2211.04292}}.

\end{thebibliography}\endgroup

\end{document}